\documentclass[3p]{elsarticle}

\usepackage{lineno,hyperref}
\usepackage{textcomp}
\modulolinenumbers[5]
\usepackage{amsmath}                        
\journal{Nuclear Instruments and Methods in Physics Research Section A}
\usepackage[section]{placeins}              

\usepackage{xcolor}

\begin{document}

\begin{frontmatter}

\title{Feeding, Reading out, and Digitizing one Channel Electronic Board for the S12572-100P Hamamatsu Photodiode}

\author{L.J.~Arceo} \corref{mycorrespondingauthor}
\cortext[mycorrespondingauthor]{Corresponding author}
\ead{miquel@fisica.ugto.mx,+52(477)7885100e8444}

\author{J.~F\'{e}lix}
\ead[url]{http://laboratoriodeparticulaselementales.blogspot.mx/}

\address{Laboratorio de Part\'{i}culas Elementales, Divisi\'{o}n de Ciencias e Ingenier\'{i}as campus Le\'{o}n, Universidad de Guanajuato, \\ 
 Loma del Bosque, 103, Col. Lomas del Campestre, C.P. 37150, Le\'{o}n, Guanajuato, M\'{e}xico.}

\begin{abstract}
The advantages of the solid state photodiodes, Hamamatsu S12572-100P type, are the reduced occupied volume, the small operation voltage, the high amplification factor, the magnetic field effects insensibleness, and other desirable physical characteristics against the conventional big vacuum-phototubes technical features. Here are the studies on a feeding, reading out, and digitizing one channel electronic board for the S12572-100P Hamamatsu photodiode. The applications are very general in cosmic ray detectors, high energy particle detectors, and radiation detection in general. The results are based on a study of 15 electronic boards (ten feeding and reading out boards and five digitizing boards) planned, designed, constructed, and tested for these purposes. The waveband of operation varies within a range from 100 Hz to 1 MHz; the ratio of outgoing signal to incoming signal (attenuation factor) is about 82\% between 100 Hz-10 kHz, is 72\% between 10 kHz-1 MHz, and is 78\% between 100 Hz-1 MHz; phase shifts between the incoming and outgoing signals are close to zero degrees for all frequencies, and occasionally have values different from zero degrees; digitizing efficiency is 99.99\%; signal transit time is about 738.08$\pm$62.38 ps; and digitizing time is about 2.88$\pm$0.15 ns. The electronic board one channel ensemble works as it was planned. More technical details and physical characteristics are described in this paper.
\end{abstract}

\begin{keyword}
Photodiodes \sep high vacuum-phototubes \sep cosmic ray detectors \sep waveband \sep cosmic radiation \sep electronic boards \sep attenuation factor \sep phase shift \sep digitizing efficiency \sep transit time \sep digitizing time.
\MSC[2000] xx-xx\sep  xx-xx
\end{keyword}

\end{frontmatter}


\section{Introduction}
Solid state photodiodes have many applications, with many advantages against the big high-vacuum-photomultipliers, in many fields of physics, like data communications, aerospace, laser range finders, radiation detection in high energy physics and cosmic ray physics, medicine, and other areas of science, due to their physical qualities like high speed, high sensitivity, big quantum efficiency, small rising time, broad wave-length range of detection, unaffected by external magnetic fields, good operation at moderate ambience temperature, small volume, small operation voltage, big amplification factor, and low cost.

Between the photodiodes based on Silicon, the S12572-100P Hamamatsu photodiode is an excellent option, due to its good physical properties and good price, on many applications including radiation detection for high energy physics and cosmic ray studies, for both research and teaching \textcolor{blue}{\cite{Datasheet-Hamamatsu}}.

Normally, Hamamatsu company only supplies a circuit hint on how to connect and feed this device, leaving to the user many possibilities on how to use the photodiode. Elsewhere we have presented some applications of this photodiode \textcolor{blue}{\cite{XXXI-ReunionAnualDPyC,ICHEP2016,SENIE2017}}. Here we present ample technical details on how to electrically feed this photodiode, on how to read out its signal, on how to digitize its signal, on the proposed electronic boards and on the experimental setup to characterize the S12572-100P Hamamatsu photodiode.

\section{Front-end electronic boards}
Basically, in this proposed experimental system, there are three separate electronic boards, planned, designed, constructed and tested to run the S12572-100P Hamamatsu photodiode: 1). Connection board, where this photodiode is placed to attach it mechanically to the detector material; 2). Feeding and reading out electronic board, to apply the feeding voltage, to create the analogical signal and to read it out; and 3). Digitizing electronic board, where the analogical signal is digitized and sent out to the data acquisition system (DAQ). All of them were separately planned, designed, and constructed to evaluate them on each stage. In future designs they will be in just one electronic board, specially, the last two boards, the feeding and reading out board and the digitizing board. Ten feeding and reading out boards and five digitizing boards were manufactured and tested. The present results are based on these electronic boards. The results were satisfactory and we operated S12572-100P Hamamatsu photodiode under good and safe conditions.  

\subsection{Connection electronic board} \label{ConnElecBoard}
\begin{figure}[ht!]
    \centering
    \includegraphics[width=468pt]{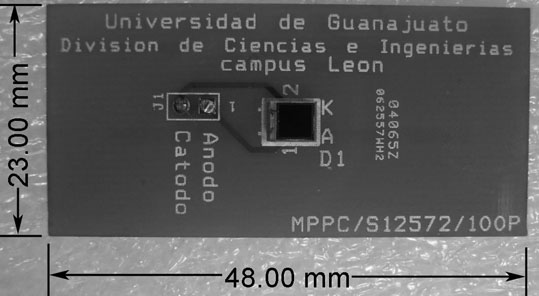}
    \caption{Connection electronic board top layer.}
    \label{FigConElecBtopLay}
\end{figure}

\begin{figure}[ht!]
    \centering
    \includegraphics[width=468pt]{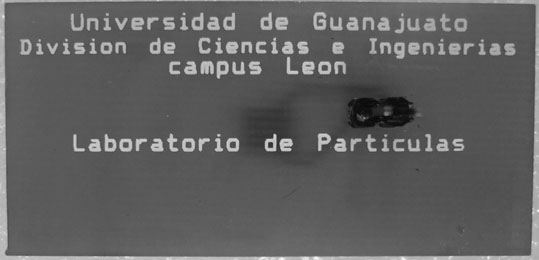}
    \caption{Connection electronic board bottom layer.}
    \label{FigConElecBbotLay}
\end{figure}

In \textcolor{blue}{Fig. \ref{FigConElecBtopLay}} and \textcolor{blue}{Fig. \ref{FigConElecBbotLay}}, there are two different views of the connection electronic board. It is very simple one. It is where the S12572-100P Hamamatsu photodiode surface mount type is soldered on the top layer; the anode and cathode terminals go to 2-pin header connector \textcolor{blue}{\cite{2-pinTerminalV}}; the \textcolor{blue}{Fig. \ref{FigConElecBbotLay}} displays this 2-pin header connector installed from the bottom layer and soldered on the top layer; and then it goes to 2-pin receptacle connector of the feeding and reading out electronic board, see \textcolor{blue}{Fig. \ref{FigAttConElecB}}. The connection electronic board is used to provide mechanical rigidity to the photodiode connection and to attach it to the detection material, in this case an 1 in $\times$ 2 in $\times$ 8 in Aluminum bar.

We proposed in our design a 90-degree relative position between connection electronic board, attached and assembled to an Aluminum bar which functions as a detector, and the feeding and reading out electronic board.

\subsection{Feeding and reading out electronic board}
\begin{figure*}[ht!]
    \centering
    \includegraphics[width=468pt]{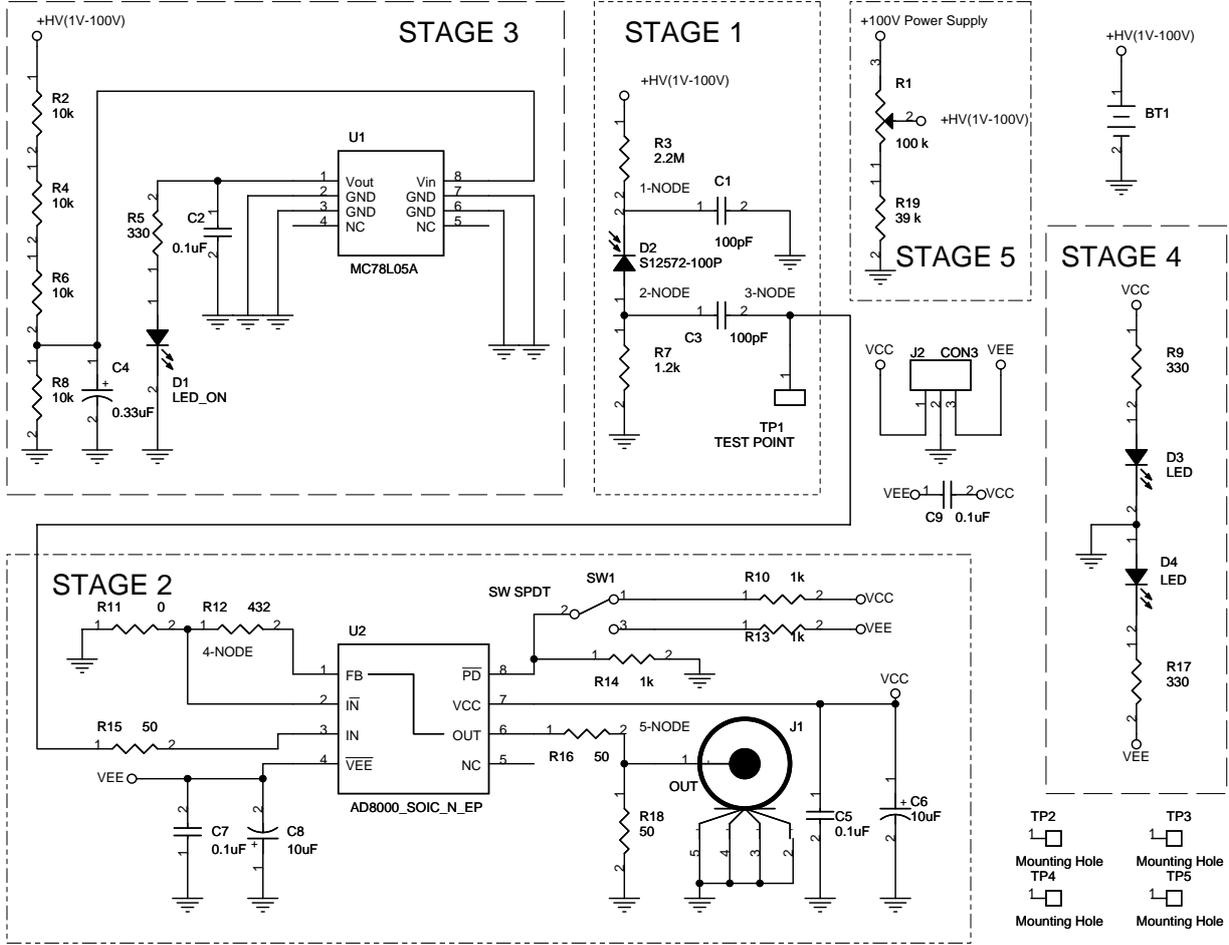}
    \caption{Schematic diagram of the feeding and reading out electronic board for the S12572-100P Hamamatsu photodiode.}
    \label{FigSchDiagFeedElecB}
\end{figure*}

In this subsection we report on a detailed technical description of the electronic boards designed, constructed, and operated to feed the S12572-100P Hamamatsu photodiode, to generate and read out its analogue signal. See \textcolor{blue}{Fig. \ref{FigSchDiagFeedElecB}}.\hfill

In \textcolor{blue}{Fig. \ref{FigSchDiagFeedElecB}} there is the schematic diagram  design of the feeding and reading out electronic board for the S12572-100P Hamamatsu photodiode.\hfill

We divided this electronic board into five stages as follows:

\begin{enumerate}[1.]
\item Feeding and reading out electronic circuit (STAGE 1).
\item Operational amplifier electronic circuit (STAGE 2).
\item Feeding and reading out operation LED signal (STAGE 3).
\item Operational amplifier operation LED signal (STAGE 4).
\item External circuit for tuning voltage (STAGE 5).
\end{enumerate}

This electronic board has two connectors to get electrical feeding. First, the 2-pin header connector (BT1) \textcolor{blue}{\cite{2-pinShroud-RA}} is used to apply voltages between +1 Vdc and +100 Vdc; it provides electrical power to STAGE 1 and STAGE 3, voltages between  1Vdc-100Vdc, with XLN10014 model, by BK Precision, power supply \textcolor{blue}{\cite{UserManual-XLN10014}}. Second, the 3-pin header connector \textcolor{blue}{\cite{3-pinShroud-RA}} (J2) is used to apply two fixed voltages: +5 Vdc, (VCC) power supply input, and -5 Vdc, (VEE) power supply input; it provides power to STAGE 2 and STAGE 4, with 72-6905 \textcolor{blue}{\cite{Datasheet-72-6905}} model, by TENMA, power supply.

The STAGE 1 is composed of three voltage nodes and one power supply:
The +HV(1Vdc-100Vdc) power supply input contains one 2.2-Mohm resistor (R3), directly connected to 1-NODE; the R3 resistor limits electrical current from power supply avoiding photodiode damage.

The 1-NODE consists of three passive components: First, avalanche photodiode (D2) (connected with a 2-pin receptacle bond \textcolor{blue}{\cite{2-pinSocketRA}}) with reverse polarity and connected to 2-NODE; second, the 1-NODE has one 100-pF capacitor (C1) connected to ground, its function is to reduce power supply noise; third, the R3 resistor is bonded to +HV(1Vdc-100Vdc) power supply input.

The 2-NODE voltage node is formed by the following components: Avalanche photodiode (D2) bonded to 1-NODE; 1.2-kohm resistor (R7) connected to ground; and 100-pF  capacitor (C3) bonded to 3-NODE. When the particle deposits its energy in the detection material and photons are created, the D2 photodiode gathers them and creates a multiplied electrical signal, collected with C3 capacitor in the order of millivolts.

The 3-NODE consists of a C3 capacitor bonded to 2-NODE, a test point connector (TP1), and a 50-ohm resistor (R15) connected to IN pin (amplifier IC pin number three) of STAGE 2, (U2) circuit. The C3 capacitor\footnote{Coupling capacitor confines dc voltages to 3-NODE.} decouples analogue signal from D2 photodiode, from feeding voltage; the test point connector (TP1) is for oscilloscope testing; the R15 resistor or R\textsubscript{S} is for configuring the operational amplifier as non-inverting one and unitary amplifier (U2).

The STAGE 2 works as a noninverting unity-gain amplifier; its actual configuration is called a voltage follower or unity-gain buffer, circuit (U2). It is used when impedance matching and circuit isolation are very important and analogical signal amplification is not important. The electronic circuit used was the AD8000 model, from Analog Devices, Inc. \textcolor{blue}{\cite{Datasheet-AD8000}}. The main features are as follows: 1.5 GHz with -3 dB bandwidth (G = +1), slew rate of 4100 V/$\mu$s, 100 mA of load current with minimal distortion, and 2 to 3.6 Mohm/pF noninverting input impedance. This STAGE was not really useful for operation but for electrical protection.

The 4-NODE is formed by R11 resistor, or R\textsubscript{G}, bonded to ground; R12 resistor, or R\textsubscript{F}, connected to pin number 1 (FB) of AD8000 circuit (U2); and the 4-NODE bonded to ($\overline{IN}$) pin number 2 of AD8000 circuit (U2).
The value of R11\footnote{R\textsubscript{G} resistor is absent in the Analog Devices, Inc. circuit.}, R12\footnote{R\textsubscript{F} resistor is 432-ohm in the Analog Devices, Inc. circuit.}, and R15\footnote{R\textsubscript{S} resistor is 50-ohm in the Analog Devices, Inc. circuit.} resistors are manufacturer's recommendation.

From the STAGE 2 electronic design, the unity-gain (G = +1) configuration was considered the main configuration; the gain of 2 (G = +2) configuration was considered for testing the amplifier, or to amplify the analogue signal amplitude from the S12572-100P Hamamatsu photodiode, if necessary. In order to comply with the PCB design software rules, the resistor R11 value was assigned to 0-ohm by default. See \textcolor{blue}{Fig. \ref{FigSchDiagFeedElecB}}. To configure the noninverting amplifier to unity-gain (G = +1), the resistor R12 is fixed to 432-ohm and resistor R11 is eliminated, its footprint is left open. To configure the noninverting amplifier to gain to 2 (G = +2), the resistor R12 is fixed to 432-ohm and R11 resistor is fixed to 432-ohm.

The output analogue signal comes from pin number 6 of AD8000 (U2), which is bonded to a 50-ohm resistor (R16); the other end of the R16 resistor is connected to 5-NODE; this voltage node contains a 50-ohm resistor (R18) bonded to ground, a SMA connector (J1) \textcolor{blue}{\cite{SMA-RA}}, and R16 resistor bonded back to the circuit (U2). The role of this node is to adjust output impedance. The final configuration was R16 resistor equals 0-ohm and R18 resistor was removed.
The switch (SW1) \textcolor{blue}{\cite{3-pinTerminalRA}} is used to turn on and turn off the AD8000 circuit (U2) (instead, the end user needs to install the shunt connector \textcolor{blue}{\cite{Shunt}} between pin 1 and 2 at SW1 to turn on AD8000 (U2), or between pin 2 and 3 to turn off).

The STAGE 3, operation LED signal, is for monitoring the correct applied voltage on STAGE 1; if the LED is on (bright red), the D2 photodiode is enabled; otherwise, disabled. The +HV(1Vdc-100Vdc) power supply input from 2-pin header connector (BT1) is connected to a voltage divider circuit formed by four 10-kohm resistors (R2, R4, R6, and R8), where R8 resistor is bonded to ground, having a quarter of +HV(1Vdc-100Vdc) power supply input, which is applied to MC78L05A circuit (U1) \textcolor{blue}{\cite{Datasheet-MC78LS05A}}; the circuit (U1) output voltage is fixed at +5 Vdc, and this is bonded to 330-ohm resistor (R5) in series with LED (D1) and then to ground; the voltage divider circuit and voltage regulator circuit (U1) allow enabling the LED (D1) signal and avoid its damage.

The STAGE 4, operation LED signal, D3 and D4, is for monitoring the proper function of STAGE 2. The STAGE 4 is formed by VCC power supply input and VEE power supply input. VCC power supply input is connected to 330-ohm resistor (R9), the other end of the R9 resistor is bonded to the anode of the LED 3 (D3), the cathode is connected to ground. VEE power supply input is connected to 330-ohm resistor (R17), the other end of the R17 resistor is bonded to the cathode of the LED 4 (D4), the anode is connected to ground.

The STAGE 5 is an external circuit for tuning voltage; it consists of 100-kohm variable resistor (R1) and 39-kohm resistor (R19) in series, the other end of the R19 resistor is bonded to ground; the end user can adjust the +HV(dc) power supply between +1 Vdc to +100 Vdc, obtaining the output voltage at pin number 2; it must be connected to 2-pin header connector (BT1) by wire. The +100 Vdc was applied using XLN10014 model BK Precision power supply. If we need to enable and feed voltage for two or more experimental setups using a single XLN10014 model BK Precision power supply, the STAGE 5 electronic circuit allows us to electric feed each experimental setup with a different feed voltage; each D2 photodiode in the experimental setup may require a different feed voltage.

In the final operation, STAGE 2 was partially disconnected removing R15, R16, and R18 resistors. 
3-NODE from STAGE 1 was bonded to SMA connector (J1), pin 1, by wire.

\begin{figure}[ht!]
    \centering
    \includegraphics[width=468pt]{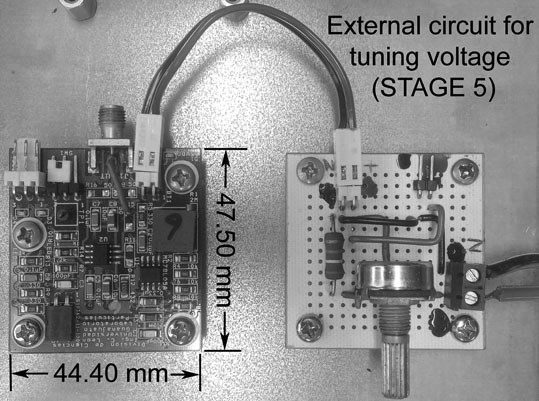}
    \caption{Manufactured feeding and reading out electronic board and its external circuit for tuning voltage (STAGE 5).}
    \label{FigManufFeedElecB}
\end{figure}

In \textcolor{blue}{Fig. \ref{FigManufFeedElecB}} it is shown an example of manufactured electronic board and its external circuit for tuning voltage (STAGE 5).

For future designs, we recommend two updates of feeding and reading out electronic board: First, replace  AD8000 Analog Device, Inc. amplifier circuit (U2) of STAGE 2 by LMH6559 Texas Instruments high-speed, closed-loop buffer \textcolor{blue}{\cite{Datasheet-LMH6559}}; it is a good option for impedance matching and electrical isolation. Second, incorporate TL783C Texas Instruments high-voltage adjustable regulator circuit \textcolor{blue}{\cite{Datasheet-TL783C}} at STAGE 5; output range has +1.25 Vdc to +125 Vdc, and 700 mA output current; the output voltage adjustment will be more precise.

\subsection{Digitizing electronic board} \label{DigElecBoard}
In this subsection we report on a detailed technical description of electronic boards designed to digitize the S12572-100P Hamamatsu photodiode analogue signal.

\begin{figure*}[ht!]
    \centering
    \includegraphics[width=468pt]{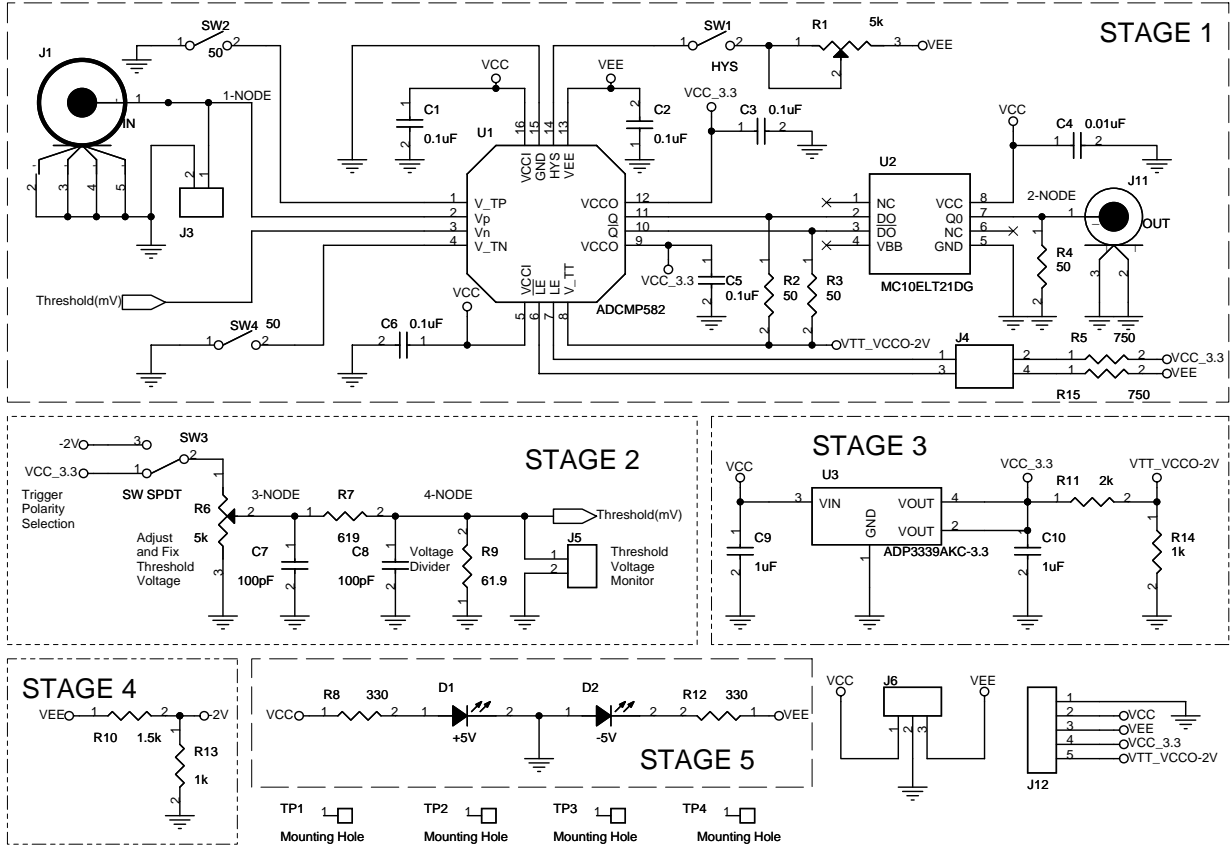}
    \caption{Schematic diagram of the digitizing electronic board.}
    \label{FigSchDiagDigitElecB}
\end{figure*}

With this electronic board, the photodiode analogue signal is compared with threshold voltage signal defined by the end user (three times bigger above noise signal was the selected criterion); if the photodiode analogical signal is bigger than the threshold voltage signal, the comparator digital output is turned on, otherwise it remains turned off. In \textcolor{blue}{Fig. \ref{FigSchDiagDigitElecB}} is shown the schematic diagram design of the digitizing electronic board. The electronic design was based on ADCMP582 chip (U1) \textcolor{blue}{\cite{Datasheet-ADCMP582}} fabricated by Analog Devices, Inc. with the following features: One-channel, ultrafast voltage comparators\footnote{180 ps propagation delay, and 100 ps minimum pulse width.}, on-chip termination at both input pins, PECL (positive emitter-coupled logic) logic family output drivers, resistor-programmable hysteresis, and differential latch control option.

The electronic board design is divided into five stages as follows:

\begin{enumerate}[1.]
\item Comparator electronic circuit (STAGE 1).
\item Threshold electronic circuit (STAGE 2).
\item VCC$\_$3.3 voltage and VTT$\_$VCCO-2V voltage electronic circuit (STAGE 3).
\item -2V voltage electronic circuit (STAGE 4).
\item Digitizing electronic board LED signal (STAGE 5).
\end{enumerate}

STAGE 1 was based on ADCMP582 (U1) and MC10ELT21DG (U2) \textcolor{blue}{\cite{Datasheet-MC10ELT21DG}} chips, the last one fabricated by ON Semiconductor\footnote{https://www.onsemi.com/}. STAGE 1 is composed of two voltage nodes: 1-NODE and 2-NODE.

1-NODE consists of a SMA connector (J1) \textcolor{blue}{\cite{SMA-RA}}, and 2-pin header connector\footnote{pin number 1 bonded to 1-NODE voltage and pin number 2 bonded to ground.} (J3) \textcolor{blue}{\cite{2-pinTerminalV}} connected to V\textsubscript{P} pin (non-inverting analogue input pin or pin number 2) of ADCMP582 (U1) chip. The function of the SMA connector (J1) is to connect electrically the feeding and reading out electronic board by SMA-to-SMA adapter \textcolor{blue}{\cite{SMApToSMAp}}, getting the analogue signal and sending it to V\textsubscript{P} pin of ADCMP582 (U1) chip. The 2-pin header connector (J3) was installed for testing (it purpose is to change the V\textsubscript{P} input impedance with an external resistor attached by end user).

2-NODE is formed by single-ended TTL digital output MC10ELT21DG (U2) chip (pin number 7), BNC connector (J11) \textcolor{blue}{\cite{BNCra}}, and the 50-ohm resistor (R4) bonded to ground. 2-NODE receives the final information of the analogue signal, and converts it into its digital version. R4 resistor was not used. 
V\textsubscript{N} pin (pin number 3) of ADCMP582 (U1) chip, inverting analogue input, is used for threshold signal input.

The comparator chip (U1) provides internal 50-ohm termination resistors for both V\textsubscript{P} and V\textsubscript{N} inputs. The V\textsubscript{TP} input pin (pin number 1), or termination resistor return pin, enables the 50-ohm termination resistor option inserting shunt connector on 2-pin header connector (SW2) \textcolor{blue}{\cite{2-pinTerminalV}}, otherwise, disables the 50-ohm termination resistor option; similarly, V\textsubscript{TN} input pin (pin number 4), or termination resistor return pin, enables the 50-ohm termination resistor option inserting shunt connector on 2-pin header connector (SW4) \textcolor{blue}{\cite{2-pinTerminalV}}, otherwise, disables the 50-ohm termination resistor option.

Differential latch input control option comes from pin number 7, or non-inverting side (LE), and from pin number 6, or inverting side ($\overline{LE}$), of ADCMP582 chip (U1). Differential latch input control option has two configurations: First, latch mode (LE = low and $\overline{LE}$ = high), the output remains as in the input state; second, compare mode (LE = high and $\overline{LE}$ = low), the output follows changes at the input state; compare mode was the selected option. The high level is VCC$\_$3.3 voltage with +3.3 Vdc from STAGE 3, and low level is VEE power supply input with -5 Vdc. Differential latch input control is formed by 2-pin dual header connector (J4) \textcolor{blue}{\cite{2-pinDualTerminalV}}, LE and $\overline{LE}$ pins are bonded to 1 and 3 pins of 2-pin dual header connector (J4) respectively; 2 and 4 pins (the opposite side 2-pin dual header connector (J4) are bonded to pull-up 750-ohm resistor (R5) in series with +3.3 Vdc voltage from STAGE 3, and bonded to pull-down 750-ohm resistor (R15) in series with VEE power supply input from 3-pin header connector (J6) \textcolor{blue}{\cite{3-pinShroud}}, respectively.
To enable compare mode, then it is necessary to insert two shunt connector \textcolor{blue}{\cite{Shunt}}: First, between 1 and 2 pins; Second, between 3 and 4 pins of 2-pin dual header connector (J4).

The V\textsubscript{TT} pin, termination return or pin number 8 of ADCMP582 chip (U1), has two goals: First, activate differential latch input control option. Second, activate PECL logic output. Analog Device Inc. recommends connect this pin to VCCO\footnote{VCCO = +3.3 Vdc, recommended voltage by Analog Device Inc.} pin (pin number 9 and 12 of U1 circuit) minus +2 Vdc; in our design, we connected V\textsubscript{TT} pin to VTT$\_$VCCO-2V voltage from STAGE 3.

The Hysteresis control, pin number 13 of ADCMP582 chip (U1), consists of 2-pin header connector (SW1) \textcolor{blue}{\cite{2-pinTerminalV}} in series with 5-kohm variable resistor (R1) bonded to VEE (-5 Vdc) power supply input. To enable hysteresis control is necessary to insert a shunt connector on 2-pin header connector (SW1), and adjust the desired amount of hysteresis; the maximum range of hysteresis that can be applied is approximately $\pm$70 mVdc. The advantage of applying hysteresis is to improve accuracy, stability, and reduction of double or multiple counting of digital signal. To obtain the advantages mentioned above, we set up the 5-kohm variable resistor (R1) at 0.9-ohm.

The two output characteristics of ADCMP582 chip (U1) are as follows: First, Q (non-inverting, pin number 11) and $\overline{Q}$ (inverting, pin number 10) pins are differential output; second, Q and $\overline{Q}$ pins are PECL logic family (positive ECL). Analog Device Inc. recommends connect Q and $\overline{Q}$ pins with a pull up 50-ohm resistor (R2 and R3, respectively) to V\textsubscript{TT}\footnote{V\textsubscript{TT} pin (pin number 8) must be bonded to VTT$\_$VCCO-2V voltage from STAGE 3.} pin of ADCMP582 chip (U1). R2 resistor, R3 resistor, and VTT$\_$VCCO-2V voltage from STAGE 3 allow to get +400 mVdc digital output at high frequencies. PECL logic family is incompatible with our module C NI‑9402 CompactRIO (cRIO) of National Instruments (NI) \textcolor{blue}{\cite{Datasheet-cRIO-9402}}. To make it compatible, this digitizing electronic board requires to use single-ended output and LVTTL logic family characteristic devices, like MC10ELT21DG chip (U2). MC10ELT21DG chip (U2) is configured, by On Semiconductor, for a differential to single-ended conversion and PECL to TTL translator; for these two characteristics, MC10ELT21DG chip (U2) is a good option for this electronic design. Q and $\overline{Q}$ pins differential output from ADCMP582 chip (U1) are bonded to D0 and $\overline{D0}$ differential input pins of MC10ELT21DG chip (U2) respectively; VCC (+5 Vdc) power supply input and ground are required; and the single-ended TTL digital output is read out using a BNC connector (J11). See 2-NODE of \textcolor{blue}{Fig. \ref{FigSchDiagFeedElecB}}.

STAGE 2 comprehends of two voltage nodes and two power supplies. 3-pin header connector (SW3) \textcolor{blue}{\cite{3-pinTerminalV}} works to select trigger polarity; pin number 1 is bonded to VCC$\_$3.3 voltage of STAGE 3; and pin number 3 is bonded to -2V voltage of STAGE 4; for positive voltage selection, shunt connector must be placed between 1 and 2 pins of 3-pin header connector (SW3); for negative voltage selection, shunt connector must be placed between 2 and 3 pins of 3-pin header connector (SW3). 5-kohm variable resistor (R6) works to adjust and fix threshold voltage, in this way: Pin number 2 of 3-pin header connector (SW3) is bonded to pin number 1 of R6 variable resistor, pin number 2 of 5-kohm variable resistor (R6) is bonded to 3-NODE (the output voltage of this pin is -2 Vdc to +3.3 Vdc), and pin number 3 of R6 variable resistor is bonded to ground.

3-NODE and 4-NODE correspond to the input signal and the output signal of the voltage divider circuit formed by 619-ohm (R7) and 61.9-ohm (R9) resistors; 100-pF (C7) and 100-pF (C8) capacitors help eliminate noise from the power supply.

3-NODE is formed by R7 resistor bonded to 4-NODE, C7 capacitor connected to ground, and pin number 2 of R6 variable resistor.

4-NODE is formed by R7 resistor bonded back to 3-NODE, C8 capacitor connected to ground, R9 resistor bonded to ground, 2-pin header connector (J5)\footnote{Pin number 1 is bonded to 4-NODE, and pin number 2 is bonded to ground.} \textcolor{blue}{\cite{2-pinTerminalV}} works to threshold voltage monitor, and V\textsubscript{IN} pin (pin number 3) of ADCMP582 chip (U1) from STAGE 1 (hierarchical port Threshold (mV) of STAGE 2 is connected to hierarchical port Threshold (mV) of STAGE 1, see \textcolor{blue}{Fig. \ref{FigSchDiagDigitElecB}}). Output voltage obtained on 2-pin header connector (J5), or threshold voltage monitor, was between -134 mVdc and +298 mVdc.

STAGE 3 was designed to generate two output voltages: VCC$\_$3.3 and VTT$\_$VCCO-2V. 

To generate VCC$\_$3.3 voltage a ADP3339AKC-3.3-RL chip (U3), of Analog Devices, Inc. \textcolor{blue}{\cite{Datasheet-ADP3339}}, was used; this chip is a positive linear regulator with Low Dropout Output (LDO), with input voltage range from +2.8 Vdc to +6 Vdc, and a fix +3.3 Vdc output; the capacitors\footnote{manufacturer recommended value.} C9 and C10 were fixed at 1 $\mu$F; pin number 3 of U3 chip is bonded to VCC power supply input; pin number 1 of U3 chip is bonded to ground; and VCC$\_$3.3 voltage was obtained from output pins number 2 and 4. 

To generate VTT$\_$VCCO-2V voltage, we used a voltage divider circuit. This circuit is formed by the following components: 1-kohm resistor (R14) connected to ground and 2-kohm resistor (R11) bonded back to VCC$\_$3.3 voltage. Output voltage obtained on R14 resistor was +1.1 Vdc.

STAGE 4 generates -2V voltage (-2 Vdc) using a voltage divider circuit, see \textcolor{blue}{Fig. \ref{FigSchDiagDigitElecB}}. VEE power supply input provides the input voltage (-5 Vdc), and the -2V voltage is the output voltage. The -2V voltage is the drop voltage on the bonded to VEE power supply input 1.5-kohm resistor (R10) and on the bonded to ground 1-kohm resistor (R13). The output voltage obtained on R13 resistor was -2 Vdc.

STAGE 5, operation LED signal, D1 and D2 is for monitoring the proper function of STAGE 1. The STAGE 5 is formed by VCC power supply input and VEE power supply input. VCC power supply input is connected to 330-ohm resistor (R8), the other end of the R8 resistor is bonded to the anode of the LED (D1), and the cathode is connected to ground. VEE power supply input is connected to 330-ohms resistor (R12), the other end of the R12 resistor is bonded to the cathode of the LED (D2), and the anode is connected to ground. If the LED (D1) and LED (D2) are on (bright red), the  digitizing electronic board is enable; otherwise, unable.

The digitizing electronic board has one connector to get voltage feeding. The 3-pin header connector (J6) is used to apply two fixed voltages: +5 Vdc or VCC power supply input, and -5 Vdc or VEE power supply input; it provides power to STAGE 1, 3, 4, and 5 with 72-6905 TENMA power supply.

The 5-pin header connector (J12) \textcolor{blue}{\cite{5-pinTerminalV}} was incorporated for test; ground is bonded to pin number 1, VCC power supply input is bonded to pin number 2, VEE power supply input is bonded to pin number 3, VCC$\_$3.3 voltage is applied to pin number 4, and VTT$\_$VCCO-2V voltage is applied to pin number 5.

\begin{figure}[ht!]
    \centering
    \includegraphics[width=450pt]{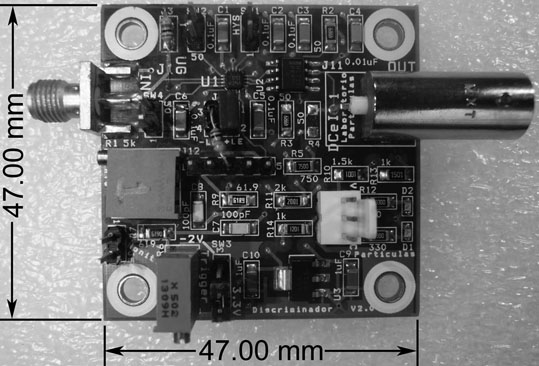}
    \caption{Manufactured digitizing electronic board.}
    \label{FigManufDiagDigitElecB}
\end{figure}

The manufactured electronic board is illustrated in \textcolor{blue}{Fig. \ref{FigManufDiagDigitElecB}}.

\section{Front-end electronic boards' characterization}
In this section we report on the procedures of evaluating the physical characteristics of the connection electronic board, feeding and reading out electronic board, and digitizing electronic board for the S12572-100P Hamamatsu photodiode \textcolor{blue}{\cite{Datasheet-Hamamatsu}}.

We divided this evaluation procedure into five subsections as follows:

\begin{enumerate}[1.]
\item S12572-100P Hamamatsu photodiode connection electronic board.
\item Feeding and reading out electronic board attenuation factor and phase shift.
\item Digitizing-electronic-board digitizing efficiency error and digitizing efficiency.
\item Feeding and reading out electronic board transit time. 
\item Digitizing electronic board digitizing time.
\end{enumerate}

\subsection{S12572-100P Hamamatsu photodiode connection electronic board}
There is no electronic evaluation of this board. See \textcolor{blue}{Fig. \ref{FigConElecBtopLay}} and \textcolor{blue}{Fig. \ref{FigConElecBbotLay}}.
Mechanically it is fine and fits all required characteristics. 

\subsection{Feeding and reading out electronic board attenuation factor and phase shift}
We describe the attenuation factor (output signal amplitude divided by input signal amplitude) of the analogical signal as function of frequency, and the phase shift (the relative phase shift of the output signal with respect to the input signal) also as a function of frequency. The attenuation factor is very important for the S12572-100P Hamamatsu photodiode calibration. 

\subsubsection{Connections}
\begin{figure}[ht!]
    \centering
     \includegraphics[width=458pt]{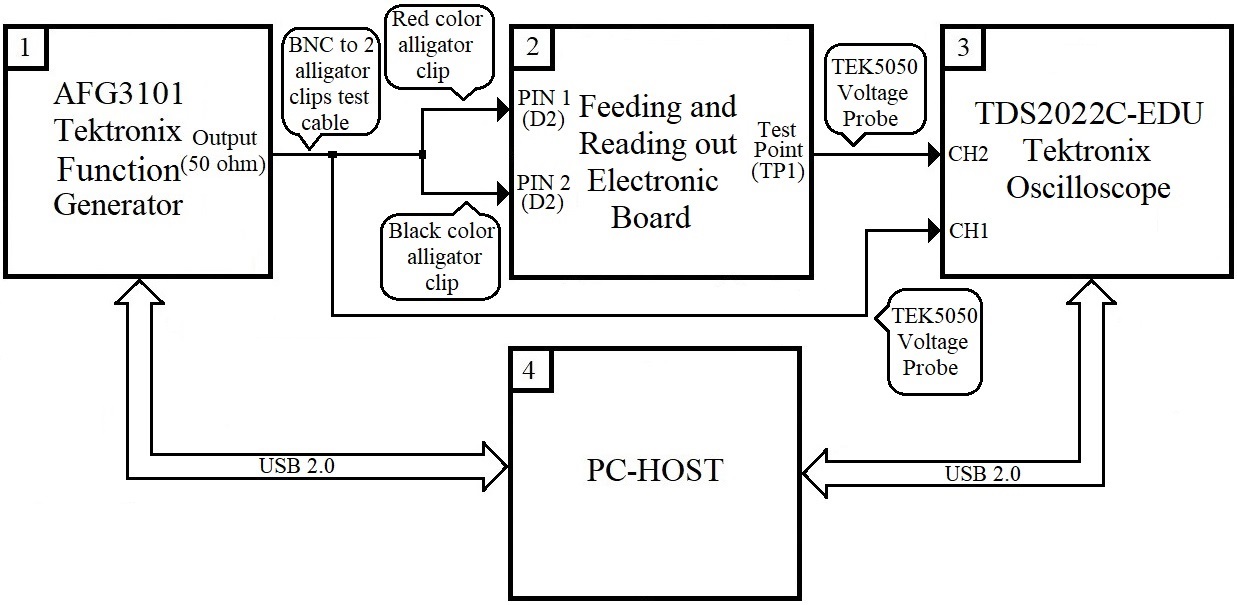}
    \caption{Block diagram of the experimental setup to measure attenuation factor and phase shift as function of frequency.}
    \label{FigBloDiaAFaPS}
\end{figure}

In  \textcolor{blue}{Fig. \ref{FigBloDiaAFaPS}} is illustrated the block diagram of attenuation factor and phase shift measurement experimental setup.

\textcolor{blue}{Fig. \ref{FigBloDiaAFaPS}} contains four blocks as follows:

\begin{enumerate}[1.]
\item BLOCK 1. AFG3101 Tektronix function generator \textcolor{blue}{\cite{UserManual-AFG3101-UM}}.
\item BLOCK 2. Feeding and reading out electronic board.
\item BLOCK 3. TDS2022C-EDU Tektronix Oscilloscope \textcolor{blue}{\cite{UserManual-TDS2022C-EDU}}.
\item BLOCK 4. PC-host (1525 Dell Inspiron) with LabVIEW\textsuperscript{TM}.
\end{enumerate}

BLOCK 1: It represents an AFG3101 Tektronix function generator we used to simulate an input analogue signal, with similar characteristics of an analogue signal from the S12572-100P Hamamatsu photodiode, with the advantage of controlling the analogue signal frequency for evaluation purposes.

BLOCK 2: It represents the feeding and reading out electronic board under characterization. It has three connectors: Input 2-pin receptacle connector (D2); output SMA connector (J1); and output test point (TP1). See \textcolor{blue}{Fig. \ref{FigSchDiagFeedElecB}}.

BLOCK 3: It represents the TDS2022C-EDU Tektronix oscilloscope. We used it to measure the analogue input signal (a V\textsubscript{Pk-Pk} of CH1), the analogue output signal (a V\textsubscript{Pk-Pk} of CH2) of the feeding and reading out electronic board, and the phase shift of CH2 with respect to CH1.

BLOCK 4: It represents the PC-host with  LabVIEW\textsuperscript{TM}\footnote{Virtual Instrument Engineering Workbench is a system-design platform and development environment for a visual programming language.} \textcolor{blue}{\cite{LabVIEW-book}} \textcolor{blue}{\cite{LabVIEW}} installed. We used this PC-host to configure and to control the AFG3101 Tektronix function generator and the TDS2022C-EDU Tektronix oscilloscope, to read out the oscilloscope measurement, and to save data text file measurements. 

The connection between AFG3101 Tektronix function generator and feeding and reading out electronic board was done using a BNC-to-alligator clip cable \textcolor{blue}{\cite{BNC-to-Alligator}}. The BNC is used to connect AFG3101 Tektronix function generator; the alligator clip, to connect feeding and reading out electronic board via 2-pin receptacle\footnote{Red color alligator clip is connected to pin number 1 (anode) and black color alligator clip to pin number 2 (cathode).} connector and a board-to-board adapter connector \textcolor{blue}{\cite{TSW-102-09-T-S-RE}}.

The connection between feeding and reading out electronic board and TDS2022-EDU Tektronix oscilloscope was done using two TPP0101 Tektronix voltage probes \textcolor{blue}{\cite{Datasheet-TPP0101}}, as follows:

First probe connection is between 2-pin receptacle connector (D2) of the feeding and reading out electronic board and oscilloscope CH1. Probe hook-tip is connected to pin number 1 (anode) and alligator clip is connected to pin number 2 (cathode) via a board-to-board adapter.

Second probe connection is between test point (TP1) of feeding and reading out electronic board and oscilloscope CH2. Probe hook-tip is connected to TP1 and alligator clip is connected to SMA ground (J1). 

The connection between AFG3101 Tektronix function generator and PC-host was done using an USB 2.0 A-to-B cable (printer cable).

The Connection between TDS2022-EDU Tektronix oscilloscope and PC-host was done using an USB 2.0 A-to-B cable (printer cable).

\subsubsection{Measurement processes}
The measurement processes consisted on seven steps - to configure, to control, to read out AFG3101 Tektronix function generator and TDS2022-EDU Tektronix oscilloscope, and to save measurements-, as follows:

\begin{enumerate}[Step 1.]
\item We turn on measurement equipment (AFG3101 Tektronix function generator, TDS2022-EDU Tektronix oscilloscope, and PC-host). 50-ohm output connector of the AFG3101 Tektronix function generator was disabled by default.
\item We open LabVIEW\textsuperscript{TM}; the getting started window appears. We use this window to select and open the AttenuationF$\_$PhaseS.vi file\footnote{The Characterization-1.zip folder contains AttenuationF$\_$PhaseS.vi LabVIEW program, and can be downloaded from the supplementary material.}, and run it by clicking on the run button. It is a home made program; we based it on the programming manual of the function generator and of the oscilloscope \textcolor{blue}{\cite{ProgramManual-AFG3101-PROGRAM,ProgramManual-TDS2022C-PROGRAM}}.
\item We set up AttenuationF$\_$PhaseS.vi parameters as follows:
\begin{enumerate}[3.1]
\item Function generator tap control. We chose and fixed positive exponential decay waveform and 0 Vdc offset. Amplitude: +2000 mVpp ( voltage peak-to-peak ); load impedance: High-Z\footnote{The High-Z, load impedance, is infinity, and its value is 9.9$\times10^{+37}$ohm \textcolor{blue}{\cite{ProgramManual-AFG3101-PROGRAM}}.}; output polarity: Off.
\item Oscilloscope tap control. CH1 Coupling: DC; CH1 bandwidth limit: Off; CH1 invert input signal: Off; CH1 attenuation: 10X; CH1 vertical scale: 500 mV/DIV; CH2 coupling: DC; CH2 bandwidth limit: Off; CH2 invert input signal: Off; CH2 attenuation: 10X; CH2 vertical scale: 500 mV/DIV; trigger source: CH1; trigger slope: Rise; trigger mode: Auto; trigger type: Edge; measure 1 menu source and type: CH1, V\textsubscript{Pk-Pk}; measure 2 menu source and type: CH2, V\textsubscript{Pk-Pk}; measure 3 menu source and type: CH2, phase; measure 4 menu source and type: CH1, none; measure 5 menu source and type: CH1, none.
\item AttenuationF$\_$PhaseS.vi recording setting LabVIEW program tap control. Start frequency: 100 Hz; end frequency: 1 MHz; first increase frequency: 100 Hz; change frequency: 10 kHz; second increase frequency: 10 kHz; name and path: Name and path of output data text file.
\end{enumerate}
\item We run AttenuationF$\_$PhaseS.vi LabVIEW program. The following actions happen:
\begin{enumerate}[4.1]
\item Starts sending all setup instructions to AFG3101 Tektronix function generator (step 3.1) and to TDS2022-EDU Tektronix oscilloscope (step 3.2), except recording setting (step 3.3).
\item Enables 50-ohm output connector of the AFG3101 Tektronix function generator.
\item Awaits eight seconds, for the stability of the output analogue signal.
\item Extracts the TDS2022-EDU Tektronix oscilloscope measurements: Measurement 1 (input analogue signal of the feeding and reading out electronic board), measurement 2 (output analogue signal of the feeding and reading out electronic board), and measurement 3 (CH2 phase shift with respect CH1).
\item Saves TDS2022-EDU Tektronix oscilloscope measurements 1, 2, and 3, and AFG3101 Tektronix function generator frequency value in a text file as follows: Frequency, a V\textsubscript{Pk-Pk} of CH1, a V\textsubscript{Pk-Pk} of CH2, and the phase shift of CH2 with respect to CH1.
\item Compares the actual frequency of the AFG3101 Tektronix function generator with the end frequency of recording setting LabVIEW program tap control; if the actual frequency of the AFG3101 Tektronix function generator is greater or equal than the end frequency of the recording setting LabVIEW program tap control, the program runs the following processes: Closes the text file, disables the output connector of the AFG3101 Tektronix function generator, and stops; on the contrary, it increases the frequency by 100 Hz or 10 kHz and continues to step 4.3.
\end{enumerate}
\item We closed AttenuationF$\_$PhaseS.vi and switched off all measurement equipment. We obtained a data text file for every feeding and reading out electronic board, a total of ten data text files, the ten data text files can be found in Characterization-1.zip supplementary material folder; each obtained data text file contains a total of 199 measurements, according to the configuration parameters of step 3.3; we created the first 100 measurements in steps of 100 Hz, starting at 100 Hz and ending at 10 kHz, to closely examine the region of frequencies between 100 Hz and 10 kHz; the remaining measurements, in steps of 10 kHz, starting at 10 kHz and ending at 1 MHz.
\item We run an attenuation factor vs frequency scripts to plot data for every feeding and reading out electronic board. From the TDS2022-EDU Tektronics oscilloscope, we consider a systematic error of 10\%; this systematic error was included in the least square technique to measure the attenuation factor as a function of the frequency. According to the TDS2022-EDU Tektronix oscilloscope user manual, the reading error is 6\%; the calibration error, 5\% (taken from the report of D. Mowbray, from the University of Sheffield \textcolor{blue}{\cite{use_of_an_oscilloscope}}).

\begin{enumerate}[6.1]
\item AT100Fs$\_$FITv3.c script to plot the first 100 measurements of every feeding and reading out electronic board. We obtained ten graphs, a graph for each feeding and reading out electronic board. The AT100Fs$\_$FITv3.c script and the ten graphs can be found in Characterization-1.zip supplementary material folder.
\item AT100Ss$\_$FITv3.c script to plot the remaining measurements of every feeding and reading out electronic board. We obtained ten graphics, a graph for each feeding and reading out electronic board. The AT100Ss$\_$FITv3.c script and the ten graphs can be found in Characterization-1.zip supplementary material folder.
\item AT199s$\_$FITv3.c script to plot all 199 measurements of every feeding and reading out electronic board. We obtained ten graphics, a graph for each feeding and reading out electronic board. The AT199s$\_$FITv3.c script and the ten graphs can be found in Characterization-1.zip supplementary material folder.
\end{enumerate}

\item We run two phase shift vs frequency scripts to plot data for every feeding and reading out electronic board.
\begin{enumerate}[7.1]
\item PhaseShift199sLV3.c script to plot all 199 measurements of every feeding and reading out electronic board; the plot scale is linear in both axes. We obtained ten graphics, one graph for each feeding and reading out electronic board. The PhaseShift199sLV3.c script and the ten graphs can be found in Characterization-1.zip supplementary material folder.
\item PhaseShift199sLog.c script to plot all 199 measurements of every feeding and reading out electronic board; the plot scale is logarithmic, in the horizontal axis and linear in the vertical axis. We obtained ten graphics, one graph for each feeding and reading out electronic board. The PhaseShift199sLog.c script and the ten graphs can be found in Characterization-1.zip supplementary material folder.
\end{enumerate}
\end{enumerate}

The attenuation factor measurement experimental setup performs a very similar and elementary measurement akin to the measurement that the professional equipment called vector network analyzer does \textcolor{blue}{\cite{VectorNetworkAnalyzers}}.

\begin{figure}[ht!]
    \centering
    \includegraphics[width=468pt]{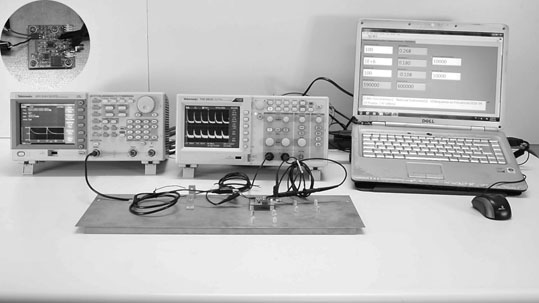}
    \caption{Experimental setup to measure attenuation factor and phase shift as function of frequency.}
    \label{FigExpSetAFaPS}
\end{figure}

In \textcolor{blue}{Fig. \ref{FigExpSetAFaPS}}, we show the attenuation factor and phase shift vs frequency measurement setup; at the top left side is displayed a figure close up that shows function generator BNC-to-alligator clip cable and oscilloscope voltage probe CH1 and CH2.

For a video of this experimental setup, of attenuation factor and phase shift measurements, \href{https://laboratoriointernacionaldeparticulaselementales.net/detecci\%C3\%B3n-de-radiaci\%C3\%B3n}{\textcolor{blue}{\underline{click here}}} or visit \textcolor{blue}{\url{https://youtu.be/zcXlBgUngn4}}

\subsection{Digitizing-electronic-board digitizing efficiency error and digitizing efficiency}
We describe the digitizing efficiency error, the digitizing efficiency, and the DAQ system used for this characterization.

The technique of digitizing efficiency error is to supply $\omega$ pulses per second (test analogue signal) at the input of the digitizing electronic board and get $\omega_{j}$ pulses per second (digital signal of test) at the output; where j is an iteration number; the digitizing  efficiency error ($eff$) is defined as follows:

\begin{equation}\label{eq:err}
eff=\frac{1}{n}\sum_{j=1}^{n}\frac{\omega_{j}-\omega}{\omega} \times 100.
\end{equation}

Here $eff$ is given in \%; n is the number of measurements; $\omega$ is the frequency; $\omega_{j}$ is the measured frequency.

The digitizing efficiency (\%) is obtained by subtracting the digitizing efficiency error (\%) from 100 percent; and it depends on the frequency.

\subsubsection{The data acquisition system}
The used DAQ system was the CompactRIO (cRIO) from National Instruments, and this system consists of three electronics components as follows: First, the cRIO-9025 controller \textcolor{blue}{\cite{UserManual-cRIO-9025}}; it is an embedded real-time controller based on core 800 MHz CPU, 512 MB DRAM, 4 GB storage; it controls and monitors the C modules across a chassis. Second, cRIO-9118 chassis \textcolor{blue}{\cite{UserManual-cRIO-9118}}; it is based on Virtex-5 LX110 FPGA technology, time-base at 40 MHz by default (80, 120, 160, or 200 MHz can be selected by software), and 8-slot to add C series modules. Third, the NI‑9402 \textcolor{blue}{\cite{Datasheet-cRIO-9402}}; it is a bidirectional four channel C series, 55 ns update rate, digital module; each channel is configurable digital I/O of the LVTTL family. 

The cRIO-9025 controller and the eight NI‑9402 C modules were installed inside the cRIO-9118 chassis obtaining 32-channel I/O DAQ system.

We used only one channel of one NI‑9402 C module, to sense digital pulses of digitizing electronic board output (J11); the measurements (counts) were sent to the PC-host by an Ethernet cable and saved them in a data text file.

To run cRIO, we create a project\footnote{NI-Paper1-DigitizingEfficiencyV1.lvproj can be found inside Characterization-2.zip supplementary material folder.} on the LabVIEW\textsuperscript{TM}, and within the project, we created two programs as follows: LabVIEW-FPGA program\footnote{DigitizingEfficiencyMeasurement-1CH.vi can be found inside Characterization-2.zip supplementary material folder.} (FPGA LabVIEW programming) to run in cRIO (to read out the digitizing electronic board measurements and transfer them into a cRIO FIFO memory) and LabVIEW VI program\footnote{PC-HOST.vi LabVIEW program can be found inside Characterization-2.zip supplementary material folder.} to run in the PC-host (to read out PC-host FIFO memory data and saved it into a data text file).

The function of these two programs are the following:
\begin{enumerate}
\item LabVIEW-FPGA program.\\
The LabVIEW-FPGA program is to activate and configure the channel number 1 (CH1 or Slot1/DIO0), and to read out the digitizing electronic board digital signal ($\omega_{j}$).\\
The LabVIEW-FPGA program contains the following four sequential circuits and one FIFO memory element \textcolor{blue}{\cite{VHDL-book}}:
\begin{enumerate}[1.]
\item One initialized rising-edge detector circuit to generate one pulse (first tick only) when the digitizing electronic board signal output changes from 0 logical to 1 logical, and to discriminate the digital signal complement of second tick go-ahead from digitizing electronic board.
\item One initialized up-counter unsigned 32-bit (U32) circuit to count one pulse (first tick only) from rising-edge detector circuit.
\item One initialized master up-counter U32 circuit to determine one-second lapse based on 40 MHz frequency clock.
\item One initialized without reset slave up-counter U32 circuit to count time, each second (the j counter from Eq. \textcolor{blue}{\eqref{eq:err}}), from the master up-counter U32 circuit.
\item One Direct Memory Access First Input First Output (DMA FIFO) memory, with these characteristics: U32 data type, target-to-host DMA type, and the number of memory elements equals 15; and with these functions: Receive data from without-reset-slave-up-counter U32 circuit (the j iteration number from Eq. \textcolor{blue}{\eqref{eq:err}}), and from up-counter U32 circuit (CH1 count or $\omega_{j}$ from digitizing electronic board, see Eq. \textcolor{blue}{\eqref{eq:err}}), and send these data to PC-host.
\end{enumerate}

The digital signal from digitizing electronic board is monitored by the rising-edge detector circuit, if one signal is detected (1 logical signal), the content of the up-counter U32 circuit increases by one (only the first tick of digital signal); if the digital signal was not detected (0 logical) or the digital signal is the complement of the first tick (1 logical in the second tick), the content of the up-counter U32 is not increased by one, and the monitoring process repeats itself indefinitely.

When the content of the master up-counter U32 circuit reaches one-second lapse ($40\times10^{6}$ ticks), the content of without reset slave up-counter U32 increases by one, it is copied to DMA FIFO (count time in seconds or j iteration number  from Eq. \textcolor{blue}{\eqref{eq:err}}); next, the last result of up-counter U32 is copied to DMA FIFO (count of CH1 or $\omega_{j}$ pulses per second, see  from Eq. \textcolor{blue}{\eqref{eq:err}}) the copy process is called interleaving-1D; next, content of up-counter U32 circuit, and content of master up-counter U32 are initialized (only one tick); the process repeats itself indefinitely. 

\item LabVIEW VI program.

The LabVIEW VI program works to collect data from cRIO. We configures LabVIEW VI program parameter as follows: Number of measurements (N) Eq. \textcolor{blue}{\eqref{eq:err}}: 1800 seconds; and data file path: Set the path of output data text file.

When we run this program, the following action happens: Actual and requested depth of DMA FIFO configure function is setup with the following two parameters: U32 data type, and number of memory elements equals 15.

The operation of the LabVIEW VI program is as follows:\hfill

LabVIEW VI program creates and opens one data text file (set data text file name at date and time from PC-host by CPU\_TIME\_V2.vi\footnote{CPU\_TIME\_V2.vi subVI program can be found inside Characterization-2.zip supplementary material folder.} LabVIEW SubVI program); when DMA FIFO memory of PC-host gets two U32 data from cRIO, then decimate 1D array function will be applied to separate data at current time in seconds (j iteration number, see Eq. \textcolor{blue}{\eqref{eq:err}}) and count of CH1 or $\omega_{j}$ of Eq. \textcolor{blue}{\eqref{eq:err}} pulses per second, and it is written to data text file; at once, the current time in seconds (j iteration number, see Eq. \textcolor{blue}{\eqref{eq:err}}) from cRIO is compared to number of measurements (N) (see Eq. \textcolor{blue}{\eqref{eq:err}}) from PC-host by a LabVIEW VI program, if its comparison is greater or equal, the LabVIEW VI program closes the data text file, and then, LabVIEW VI program shutdowns itself; otherwise, LabVIEW VI program continues.
\end{enumerate}

\subsubsection{Connections} \label{EFFconnection}
To perform the digitizing efficiency error characterization, we divided the experimental setup into two particular experimental setups, for comparison purposes, as follows: The first one is to measure the digitizing efficiency error of digitizing electronic boards; the second one, to measure the digitizing efficiency error of the AFG3101 Tektronix function generator. We define the cRIO system as the reference to measure the digitizing efficiency error of digitizing electronic boards and to measure the digitizing efficiency error of the AFG3101 Tektronix function generator. 
\begin{enumerate}[1.]
\item Experimental setup to measure the digitizing efficiency error of digitizing electronic boards.
We carried out this measurement process for each digitizing electronic board, with a total of five.

In \textcolor{blue}{Fig. \ref{FigBloDiaEFF}}, we show the block diagram of the experimental setup to measure the digitizing efficiency error of digitizing electronic boards.

\begin{figure}[ht!]
    \centering
     \includegraphics[width=468pt]{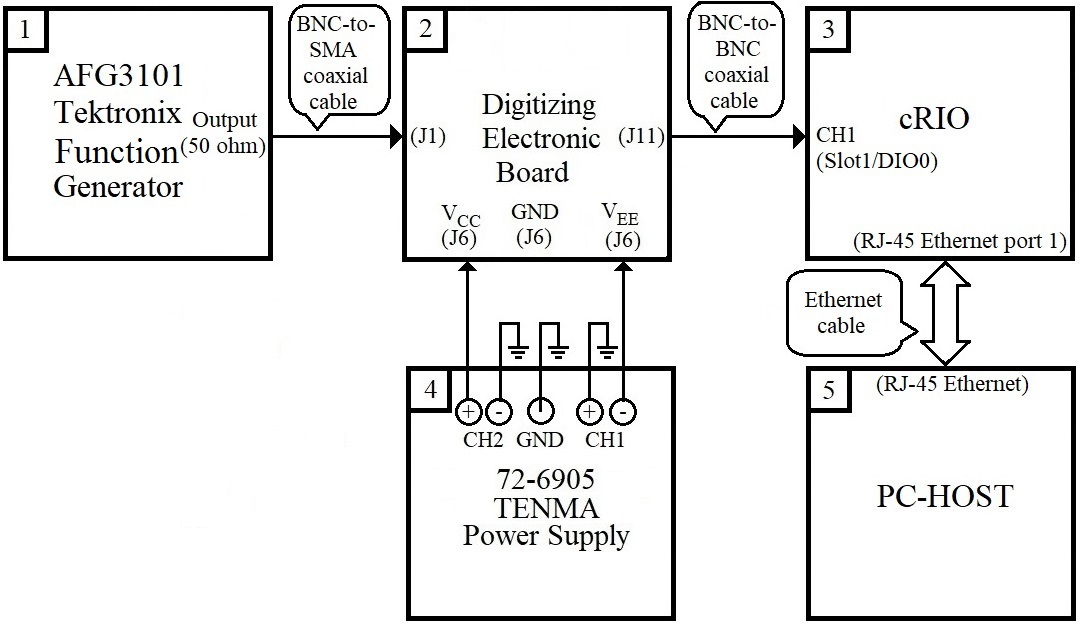}
    \caption{Block diagram of the experimental setup to measure digitizing efficiency error of digitizing electronic boards.}
    \label{FigBloDiaEFF}
\end{figure}

\textcolor{blue}{Fig. \ref{FigBloDiaEFF}} contains five blocks as follows: 

\begin{enumerate}[1.]
\item BLOCK 1. AFG3101 Tektronix function generator.
\item BLOCK 2. Digitizing electronic board.
\item BLOCK 3. DAQ system with National Instruments cRIO.
\item BLOCK 4. 72-6905 TENMA power supply.
\item BLOCK 5. PC-host (1525 Dell Inspiron) with LabVIEW\textsuperscript{TM}.
\end{enumerate}

BLOCK 1: It represents an AFG3101 Tektronix function generator. We used it to supply $\omega$ pulses per second (test analogue signal) at the input of the digitizing electronic board.\hfill

BLOCK 2: It represents the digitizing electronic board under characterization. It has one input, the SMA connector (J1), and one output, the BNC connector (J11). See \textcolor{blue}{Fig. \ref{FigSchDiagDigitElecB}}, for schematic diagram of digitizing electronic board; and see \textcolor{blue}{Fig. \ref{FigManufDiagDigitElecB}}, for manufactured of digitizing electronic board.

BLOCK 3: It represents the cRIO. We used it to count the number of pulses from digitizing electronic board per second ($\omega_{j}$). See Eq. \textcolor{blue}{\eqref{eq:err}}. 

BLOCK 4: It represents the 72-6905 TENMA power supply. We used it to generate two output voltages with respect to ground: +5 Vdc (VCC) power supply input and -5 Vdc (VEE) power supply input, to feed the digitizing electronic board. See the schematic diagram of digitizing electronic board, \textcolor{blue}{Fig. \ref{FigSchDiagDigitElecB}}.

BLOCK 5: It represents the PC-host with installed LabVIEW\textsuperscript{TM} \textcolor{blue}{\cite{LabVIEW}}. We used this PC-host to configure, to read out the Mod1/DIO0 channel of cRIO, and to save data text file measurements.\hfill

We used a BNC-to-SMA coaxial cable \textcolor{blue}{\cite{BNC-to-SMA}} to connect the AFG3101 Tektronix function generator with the digitizing electronic board. The BNC goes to the AFG3101 Tektronix function; and the SMA goes to the digitizing electronic board.

We used a BNC-to-BNC coaxial cable \textcolor{blue}{\cite{BNC-to-BNC}} to connect the digitizing electronic board with the cRIO. The BNC goes to BNC (J11) of digitizing electronic board; and the BNC goes to number 0 of slot 1 (Slot1/DIO0) BNC connector of cRIO.

We used a power cable to connect the digitizing electronic board with the 72-6905 TENMA power supply, with the following characteristic: Multiconductor unshielded cable, 26 AWG, three conductors, black jacket (ground), red jacket (VCC power supply input), and white jacket (VEE power supply input), 300 Vdc voltage rating; the ended stripped wires were connected to the 72-6905 TENMA power supply side; and 3-pin crimp housing receptacle connector \textcolor{blue}{\cite{3-pinRectacleCrimp}}, to 3-pin header connector (J6) of the digitizing electronic board side.

We used an Ethernet standard Category 5 (CAT-5) cable to connect the cRIO with PC-host; the RJ-45 Ethernet port number 1 was used to configure, program, and run the cRIO. 
\item Experimental setup to measure the digitizing efficiency error of AFG3101 Tektronix function generator.

We compared this measurement with the digitizing efficiency error of the five digitizing electronic boards; we carried out this measurement just one time.
In \textcolor{blue}{Fig. \ref{FigBloDiaRef}}, we show the block diagram of the experimental setup to measure the digitizing efficiency error of the AFG3101 Tektronix function generator.

\begin{figure}[ht!]
    \centering
    \includegraphics[width=468pt]{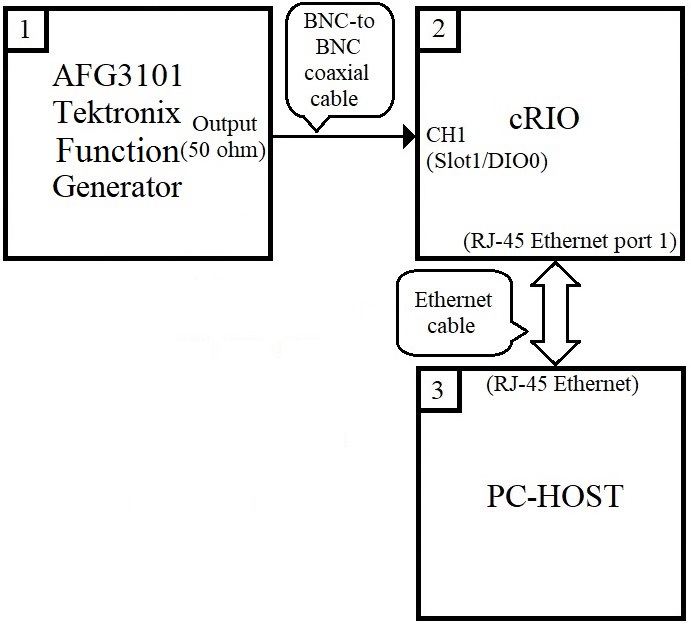}
    \caption{Block diagram of the experimental setup to measure digitizing efficiency error of the AFG3101 Tektronix function generator.}
    \label{FigBloDiaRef}
\end{figure}

From \textcolor{blue}{Fig. \ref{FigBloDiaEFF}} block diagram of the experimental setup to measure digitizing efficiency error of the digitizing electronic board, we removed the BLOCK 2 and BLOCK 4 and obtained the block diagram of the experimental setup to measure digitizing efficiency error of the AFG3101 Tektronix function generator. See \textcolor{blue}{Fig. \ref{FigBloDiaRef}}.

\textcolor{blue}{Fig. \ref{FigBloDiaRef}} contains three blocks as follows:
 
\begin{enumerate}[1.]
\item BLOCK 1. AFG3101 Tektronix function generator.
\item BLOCK 2. DAQ system with National Instruments cRIO.
\item BLOCK 3. PC-host (1525 Dell Inspiron) with LabVIEW\textsuperscript{TM}.
\end{enumerate}
 
We used a BNC-to-BNC coaxial cable to connect the AFG3101 Tektronix function generator with cRIO; the 50-ohm output of AFG3101 Tektronix function generator to input BNC connector number 0 of slot 1 (Slot1/DIO0) of cRIO.

The connection between cRIO and PC-host is unchanged from digitizing efficiency error measurement experimental setup.
\end{enumerate}

\subsubsection{Measurement processes}
There are two types of measurements, on which we are interested, as follows: The first one is the measurement of the digitizing efficiency error of the digitizing electronic boards; the second one, the measurement of the digitizing efficiency error of AFG3101 Tektronix function generator.

\begin{enumerate}[ 1.]
\item To measure the digitizing efficiency error of the digitizing electronic boards we proceed in ten steps:
\begin{enumerate}[Step 1.]
\item We assembled the experimental setup represented in \textcolor{blue}{Fig. \ref{FigBloDiaEFF}} as a block diagram.

\item We turned on 72-6905 TENMA power supply; adjust the VCC power supply input at +5 Vdc, and VEE at -5 Vdc.

\item We turned on the measurement equipment (AFG3101 Tektronix function generator, cRIO, and PC-host). 50-ohm output connector of the AFG3101 Tektronix function generator was disabled by default.

\item We set up AFG3101 Tektronix function generator and digitizing electronic board parameters as follows:

\begin{enumerate}[4.1]
\item AFG3101 Tektronix function generator: Waveform, exponential decay; amplitude, +360 mV\textsubscript{Pk-Pk} ( millivolts peak-to-peak ); load impedance, high-Z; output polarity, off; voltage offset, 0 Vdc; frequency, 1 Hz; and phase shift, 0.0$^\circ$.
\item Digitizing electronic board: Select trigger polarity (SW3), positive voltage (shunt connector must be placed between 1 and 2 pins of 3-pin header connector (SW3)); threshold voltage, +90 mVdc (in subsection \textcolor{blue}{\ref{AdFixThresProcedure}}, we will explain the adjustment procedure to setup the digitizing electronic boards and how to fix the threshold voltage of the digitizing electronic boards); hysteresis option (SW1), on (shunt connector must be placed into hysteresis option (SW1)); hysteresis control variable resistor (R1), 0.9-ohm; internal 50-ohm input impedance option (SW2 and SW4), off (shunt connectors must be release from internal 50-ohm input impedance option (SW2 and SW4)); differential latch input control option (J4), latch mode (first shunt connector must be placed between 1 and 2 pins of differential latch input control option (J4), and second shunt connector must be placed between 3 and 4 pins of differential latch input control option (J4)); and install 10-kohm resistor on the 2-pin header connector (J3).
\end{enumerate}

\item We enabled the 50-ohm output connector of the AFG3101 Tektronix function generator.

\item We launched the LabVIEW\textsuperscript{TM} and opened our project NI-Paper1-DigitizingEfficiencyV1.lvproj; in the project explorer window, expanded the plus sign located near My Computer icon, and double-clicked to PC-HOST.vi LabVIEW program. The Characterization-2.zip folder contains NI-Paper1-DigitizingEfficiencyV1.lvproj, and PC-HOST.vi LabVIEW program, and can be downloaded from the supplementary material.

\item We set up PC-HOST.vi LabVIEW program parameters as follows: Number of measurements (n) in 1800 seconds (see Eq. \textcolor{blue}{\eqref{eq:err}}); data text files path -the name of the data text file is created automatically by CPU\_TIME\_V2.vi subVI program -.

\item We run PC-HOST.vi LabVIEW program by clicking on the run button. The following actions happen: PC-HOST.vi LabVIEW program collects data at 1800 samples per each frequency value; we started the frequency at 1 Hz. PC-HOST.vi LabVIEW program shutdowns itself after 1800 seconds of runtime.

\item We increased the frequency (1 Hz) by a factor of 10 in each step; we run PC-HOST.vi LabVIEW program in each frequency step, and PC-HOST.vi LabVIEW program shutdowns itself after 1800 seconds of each step. We repeat this process (nine steps) until we reach the frequency of 1 MHz.\hfill
We proceed in the above way for each of the five digitizing electronic boards.\hfill

\item We disabled the 50-ohm output connector of the AFG3101 Tektronix function generator; we disconnected the BNC-to-SMA coaxial cable from the digitizing electronic board; we turned off the 72-6905 TENMA power supply; we released the 3-pin crimp housing receptacle connector of the power cable from the 3-pin header connector (J6) of the digitizing electronic board side; we replaced the digitizing electronic board with the next digitizing electronic board; we reconnected the 3-pin crimp housing receptacle connector of the power cable into the 3-pin header connector (J6) of the digitizing electronic board; we turned on the 72-6905 TENMA power supply; we enabled the 50-ohm output connector of the AFG3101 Tektronix function generator.
\end{enumerate}

We repeated three of above steps, for each digitizing electronic board: Eight, nine, and ten.

After the fifth digitizing electronic board, we disabled the 50-ohm output connector of the AFG3101 Tektronix function generator, we turned off the 72-6905 TENMA power supply, we disconnect the 72-6905 TENMA power supply, we disconnected the digitizing electronic board. The process of collecting data for digitizing efficiency error of digitizing electronic boards comes to end.

We obtained seven data text files for each digitizing electronic board. The Characterization-2.zip supplementary material folder contains 35 data text files for five digitizing electronics boards.

\item  To measure the digitizing efficiency error of AFG3101 Tektronix function generator we proceed in eight steps as follows:

\begin{enumerate}[Step 1.]
\item We assembled the experimental setup represented in \textcolor{blue}{Fig. \ref{FigBloDiaRef}} as a block diagram.

\item We turned on the measurement equipment (AFG3101 Tektronix function generator, cRIO, and PC-host). 50-ohm output connector of the AFG3101 Tektronix function generator was disabled by default.

\item We set up AFG3101 Tektronix function generator parameters as follows: Set waveform, pulse; set amplitude, +3 V\textsubscript{Pk-Pk} ( voltage peak-to-peak ); set load impedance, high-Z; set duty cycle, 5\%; set the voltage offset, +1.5 Vdc; set initial frequency, 1 Hz; and set phase shift, 0.0$^\circ$.

\item We enabled the 50-ohm output connector of the AFG3101 Tektronix function generator.
\item We launched the LabVIEW\textsuperscript{TM} and opened our project NI-Paper1-DigitizingEfficiencyV1.lvproj, and double-clicked to PC-HOST.vi LabVIEW program.

\item We set up PC-HOST.vi LabVIEW program parameters as follows: Set number of measurements (n) at 1800 seconds of runtime (see Eq. \textcolor{blue}{\eqref{eq:err}}), and set the data text files path.

\item We run PC-HOST.vi LabVIEW program by clicking on the run button. The following actions happen: PC-HOST.vi LabVIEW program collects data at 1800 samples per each frequency value; we started the frequency at 1 Hz. PC-HOST.vi LabVIEW program shutdowns itself after 1800 seconds of runtime.

\item We increased the frequency (1 Hz) by a factor of 10 Hz in each step; we run PC-HOST.vi LabVIEW program in each frequency step, and PC-HOST.vi LabVIEW program shutdowns itself after 1800 seconds of runtime on each step. We repeat this process (eight steps) until we reach the frequency of 1 MHz.
\end{enumerate}

After PC-HOST.vi LabVIEW program shutdowns itself, we performed the following steps: We disabled the 50-ohm output connector of the AFG3101 Tektronix function generator; we turned off AFG3101 Tektronix function generator, cRIO, and PC-host; and we disconnected all the measuring equipment. The process of collecting data for digitizing efficiency error of AFG3101 Tektronix function generator came to end.

We obtained only seven data text files for AFG3101 Tektronix function generator. The Characterization2.zip supplementary material folder contains seven data text files for digitizing efficiency error of AFG3101 Tektronix function generator.

\item We declared 42 data text file (of digitizing efficiency error, 35 data text files of five digitizing electronics boards; and digitizing efficiency error, seven data text files of AFG3101 Tektronix function generator); and we run a script, eff.c, to plot digitizing efficiency error and digitizing efficiency of five electronic boards and one AFG3101 Tektronix function generator. We obtained two graphs, a graph for digitizing efficiency error, and a graph for digitizing efficiency. The Characterization-2.zip folder contains eff.c script, two graphs, and can be downloaded from the supplementary material for close inspection.
\end{enumerate}

\begin{figure}[ht!]
    \centering
    \includegraphics[width=468pt]{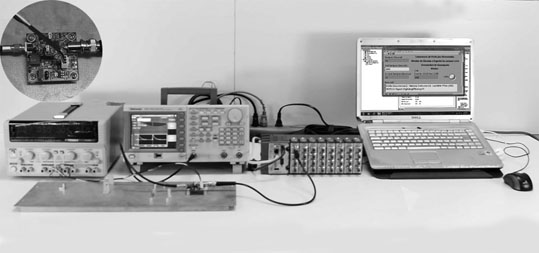}
    \caption{Experimental setup to measure digitizing efficiency error of the digitizing electronic boards.}
    \label{FigExpSetEFF}
\end{figure}

In \textcolor{blue}{Fig. \ref{FigExpSetEFF}}, we show the experimental setup to measure the digitizing efficiency error of digitizing electronic boards assembled; at the top left side is displayed a figure close up which shows power supply, power cable, AFG3101 Tektronix function generator BNC-to-SMA coaxial cable, and cRIO BNC-to-BNC coaxial cable.

\begin{figure}[ht!]
    \centering
    \includegraphics[width=468pt]{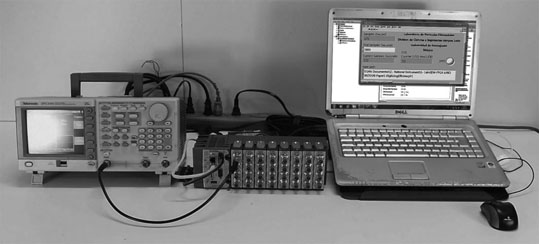}
    \caption{Experimental setup to measure digitizing efficiency error of the AFG3101 Tektronix function generator.}
    \label{FigExpSetRef}
\end{figure}

In \textcolor{blue}{Fig. \ref{FigExpSetRef}}, we show the experimental setup to measure the digitizing efficiency error of AFG3101 Tektronix function generator. 

For a video of this experimental setup, of digitizing-electronic-board digitizing efficiency error and digitizing efficiency, please \href{https://laboratoriointernacionaldeparticulaselementales.net/detecci\%C3\%B3n-de-radiaci\%C3\%B3n}{\textcolor{blue}{\underline{click here}}} or visit \textcolor{blue}{\url{https://youtu.be/eYnpFRaQVbs}}

\subsubsection{Adjust and fix threshold voltage procedure of digitizing electronic board} \label{AdFixThresProcedure}
The digitizing electronic board has one connector to read out threshold voltage, 2-pin header connector (J5) \textcolor{blue}{\cite{2-pinTerminalV}}, or threshold voltage monitor. PIN 1 is to read out threshold voltage; PIN 2, is the ground or reference. See \textcolor{blue}{Fig. \ref{FigSchDiagDigitElecB}} STAGE-2. The connection between the multimeter and the threshold voltage monitor (J5) was done using two-alligator clips to 2-pin receptacle crimp housing \textcolor{blue}{\cite{ISSM-02}} cable \footnote{The cable is homemade.}; the red color alligator clip is bonded to PIN 1; the black one, to PIN 2.

The red multimeter probe is connected to the red color alligator clip; the black one, to the black color alligator clip; The PIN 1 and PIN 2 of the homemade cable must be plugged into PIN 1 and PIN 2 of threshold voltage monitor (J5), respectively. 

We configured the multimeter in millivolts (mV). To adjust the threshold voltage, we used a small screwdriver on  R6 variable resistor to modify the potential difference on (R6), and saw the display of the multimeter at the same time.

\begin{figure}[ht!]
    \centering
  \includegraphics[width=400pt]{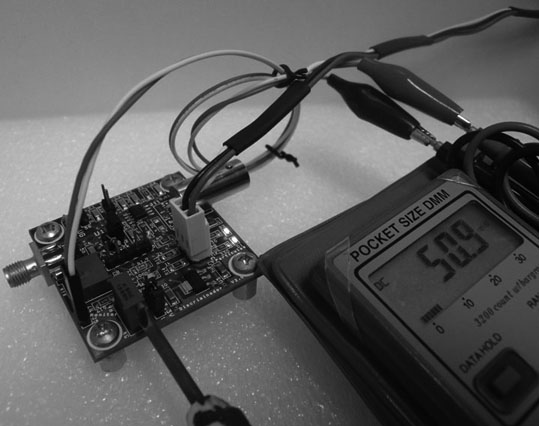}
    \caption{Threshold voltage adjustment connection.}
    \label{FigThresVolt}
\end{figure}

\textcolor{blue}{Fig. \ref{FigThresVolt}} shows the threshold voltage adjustment connection.

After fixing the threshold voltage of the digitizing electronic board, we disconnected the two-alligator clips to 2-pin receptacle crimp housing cable and the multimeter.

\subsection{Feeding and reading out electronic board transit time} \label{TTimeC}
We analyzed and measured the time an electrical signal takes to travel through the feeding and reading out electronic boards. This time interval is known as propagation delay and response time \textcolor{blue}{\cite{OPAMP-Coughlin}}. We adumbrated this process as follows: 1) connections, we enumerate and sketch all the used equipment; 2) measurement processes, we explain the sequence in which we ran the experimental setup. 

\subsubsection{Connections}
\begin{figure}[ht!]
    \centering
    \includegraphics[width=468pt]{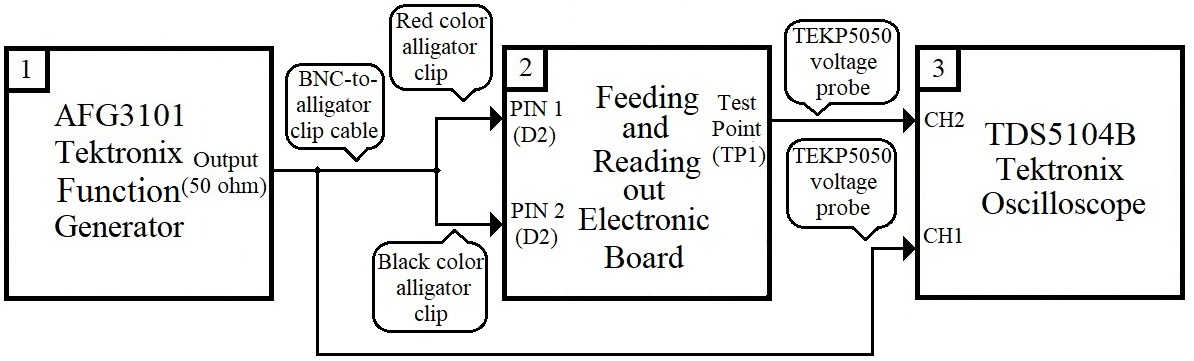}
    \caption{Block diagram of the experimental setup to measure transit time of the feeding and reading out electronic board.}
    \label{FigBloDiaTTime}
\end{figure}

In \textcolor{blue}{Fig. \ref{FigBloDiaTTime}}, we show the block diagram of the feeding and reading out electronic board transit time experimental setup.

It contains three blocks as follows:

\begin{enumerate}[1.]
\item BLOCK 1. AFG3101 Tektronix function generator.
\item BLOCK 2. Feeding and reading out electronic board.
\item BLOCK 3. TDS5104B Tektronix oscilloscope \textcolor{blue}{\cite{UserManual-TDS5104B-M}}.
\end{enumerate}

BLOCK 1: It represents an AFG3101 Tektronix function generator. We used it to supply test analogue signal at the input of the feeding and reading out electronic board.

BLOCK 2: It represents the feeding and reading out electronic board under characterization. It has one input, the 2-pin receptacle connector (D2); and two outputs, the test point (TP1), and SMA connector (J11). See \textcolor{blue}{Fig. \ref{FigSchDiagFeedElecB}}, for schematic diagram of feeding and reading out electronic board; and see \textcolor{blue}{Fig. \ref{FigManufFeedElecB}}, for manufactured feeding and reading out electronic board.

BLOCK 3: It represents the TDS5104B Tektronix oscilloscope. We used it to measure the delay of the analogue output signal (CH2) with respect to the analogue input signal (CH1) of the feeding and reading out electronic board.

We used a BNC-to-alligator clip cable to connect the AFG3101 Tektronix function generator with the feeding and reading out electronic board. The BNC goes to the AFG3101 Tektronix functions generator; the alligator clip end\footnote{Red alligator clip connected to pin number 1 (anode) and black alligator clip connected to pin number 2 (cathode).}, to 2-pin receptacle connector (D2) \textcolor{blue}{\cite{2-pinSocketRA}} using a board-to-board adapter connector \textcolor{blue}{\cite{TSW-102-09-T-S-RE}}.

We used two TEKP5050 Tektronix voltage probes \textcolor{blue}{\cite{UserManual-P5050}} to connect the feeding and reading out electronic board with the TDS5104B Tektronix oscilloscope, as follows:

Connection 1, between 2-pin receptacle connector (D2) of feeding and reading out electronic board and CH1 of the TDS5104B Tektronix oscilloscope. Probe hook-tip goes to pin number 1 (anode); and alligator clip, to pin number 2 (cathode) via a board-to-board adapter connector. The CH1 of the TDS5104B Tektronix oscilloscope is to measure the input analogue signal in the feeding and reading out electronic board.

Connection 2, between test point (TP1) of the feeding and reading out electronic board and CH2 of TDS5104B Tektronix oscilloscope. Probe hook-tip goes to TP1 connector; and alligator clip, to SMA metal (J1). This CH2 of the TDS5104B Tektronix oscilloscope is to measure the output analogue signal in the feeding and reading out electronic board.

\subsubsection{Measurement processes}
The measurement processes consisted on ten steps as follows:
\begin{enumerate}[Step 1.]
\item We assembled the experimental setup represented in \textcolor{blue}{Fig. \ref{FigBloDiaTTime}} as a block diagram.

\item We turned on the measurement equipment (AFG3101 Tektronix function generator, and TDS5104B Tektronix oscilloscope). 50-ohm output connector of the AFG3101 Tektronix function generator was disabled by default.

\item We set up AFG3101 Tektronix function generator and TDS5104B Tektronix oscilloscope parameters as follows:
\begin{enumerate}[3.1]
\item We set up AFG3101 Tektronix function generator parameters as follows: Waveform: Exponential decay; amplitude: +2 V\textsubscript{Pk-Pk} ( voltage peak-to-peak ); load impedance: High-Z; output polarity: Off; voltage offset: 0 Vdc; frequency: 1 MHz; and phase shift: 0.0$^\circ$.
\item We enabled the 50-ohm output connector of the AFG3101 Tektronix function generator.
\item We set up TDS5104B Tektronix oscilloscope parameters as follows: CH1 termination: 1-Mohm; CH1 coupling: DC; CH1 invert input signal: Off; CH1 bandwidth: Full; CH1 attenuation: 10X; CH1 vertical scale: 200 mV/DIV; CH1 deskew time: 0.0 s; CH2 termination: 1-Mohm; CH2 coupling: DC; CH2 invert input signal: Off; CH2 bandwidth: Full; CH2 attenuation: 10X; CH2 vertical scale: 200 mV/DIV; CH2 deskew time: 0.0 s; horizontal scale: 2 ns/DIV; trigger source: CH1; trigger slope: Rise; trigger mode: Auto; trigger type: Edge; measure 1 menu source and type: CH1, Delay (C1, C2). The Delay (C1, C2) measurement procedure is from C1 to C2 (in time units).
\item We set up the TDS5104B Tektronix oscilloscope front-panel print button as a save button quick storage of screenshots as follows: From the menu bar of the TDS5104B Tektronix oscilloscope, we selected the Save As..., dialog boxes will be displayed, and we fill the next parameters as follows: Base file name: Defined by us; count: 000; save as type: JPEG; auto-increment file name check box: On. Click the Save button.
\end{enumerate}
\item We saved at least 30 screenshots of a waveform by clicking 30 times on the front-panel print button.
\item We disabled the 50-ohm output connector of the AFG3101 Tektronix function generator; we disconnected the BNC-to-alligator clip cable and two TEKP5050 Tektronix voltage probes from the board-to-board adapter connector; we disconnected the board-to-board adapter connector from the feeding and reading out electronic board; we replaced the feeding and reading out electronic board with the next feeding and reading out electronic board; we connected again the board-to-board adapter connector into the 2-pin receptacle connector (D2) of the feeding and reading out electronic board; we connected the BNC-to-alligator clip cable and two TEKP5050 Tektronix voltage probes on the board-to-board adapter connector; we enabled the 50-ohm output connector of the AFG3101 Tektronix function generator.
\item We repeated the step four, and the step five for each feeding and reading out electronic board. After the tenth feeding and reading out electronic board, we disabled the 50-ohm output connector of the AFG3101 Tektronix function generator, we turned off the AFG3101 Tektronix function generator, and TDS5104B Tektronix oscilloscope, we disconnected the tenth feeding and reading out electronic board, AFG3101 Tektronix function generator, and TDS5104B Tektronix oscilloscope. The process of collecting data for transit time of feeding and reading out electronic boards comes to end. The Characterization-3.zip supplementary material folder contains at least 30 screenshots of a waveform of each feeding and reading out electronic board; the delay measurement or Delay (C1, C2) is displayed.

\item We read the Delay (C1, C2) measurement from each screenshot of a waveform, and we wrote into a data text file. This process was not automatic.

\item We repeated the above step seven for each feeding and reading out electronic board screenshot of a waveform. We obtained one data text file for each feeding and reading out electronic board. The Characterization-3.zip supplementary material folder contains ten data text files for ten feeding and reading out electronics boards.

\item We read ten feeding and reading out electronic board data text files and compute the average values and standard deviations of these ten data text files, using Statistics02 TTime.vi LabVIEW program, contained in the Characterization-3.zip supplementary material folder.

\item We wrote ten statistical error results from Statistics02 TTime.vi LabVIEW program into the script PlotErrorsTTimeNI-4.c; we run a script, PlotErrorsTTimeNI-4.c, we obtained one plot of the ten feeding and reading out electronic boards transit time statistical error. The Characterization-3.zip supplementary material folder contains the PlotErrorsTTimeNI-4.c script, and one plot.
\end{enumerate}

\begin{figure}[ht!]
\centering
\includegraphics[width=468pt]{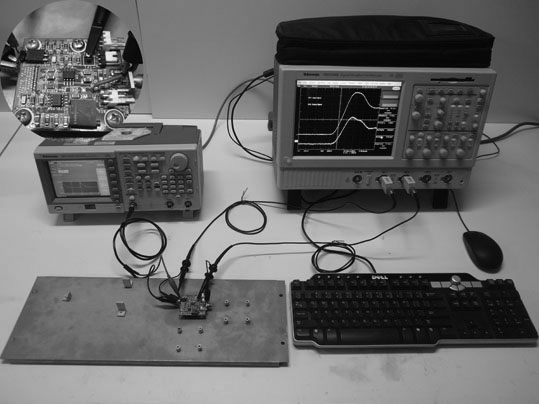}
\caption{Experimental setup to measure transit time of the feeding and reading out electronic board.}
\label{FigExpSetTTime}
\end{figure}

In \textcolor{blue}{Fig. \ref{FigExpSetTTime}}, we show the experimental setup to measure the transit time of feeding and reading out electronic assembled board; at the top left side is displayed a figure close up that shows function generator BNC-to-alligator clip cable, and oscilloscope voltage probe CH1 and CH2.

\newpage
\subsection{Digitizing electronic board digitizing time} \label{DTimeC}
Digitizing electronic board is a basic analogue-to-digital converter (ADC) with the following parameters: 1-bit resolution, high-speed, single analogue input, single analogue output, and basic control signal for microprocessor connection such as: Differential latch control option (LE, $\overline{LE}$) \textcolor{blue}{\cite{Datasheet-ADCMP582}}. The digitizing electronic, which satisfies LVTTL family logic interface and high speed requirements, is an excellent option to be connected to the National Instruments cRIO.
 
We want to know and measure the time that the digitizing electronic board spends to convert the analogue signal into the digital one; this time interval is known as conversion time \textcolor{blue}{\cite{OPAMP-Coughlin,EES-Storey}} or digitizing time. The digitizing time of the digitizing electronic board is limited by the transit time of comparator circuit, ADCMP582 chip (U1), and logic gate circuit, MC10ELT21DG (U2) \textcolor{blue}{\cite{Datasheet-MC10ELT21DG}}. In \textcolor{blue}{Fig. \ref{FigSchDiagDigitElecB}} we have shown the schematic diagram design of the digitizing electronic board.

We adumbrated this process as follows: 1) connections, we enumerate and sketch all the used equipment; 2) measurement processes, we explain the sequence in which we ran the experimental setup. 

\subsubsection{Connections} \label{DTimeConnection}
\begin{figure}[ht!]
\centering  
\includegraphics[width=468pt]{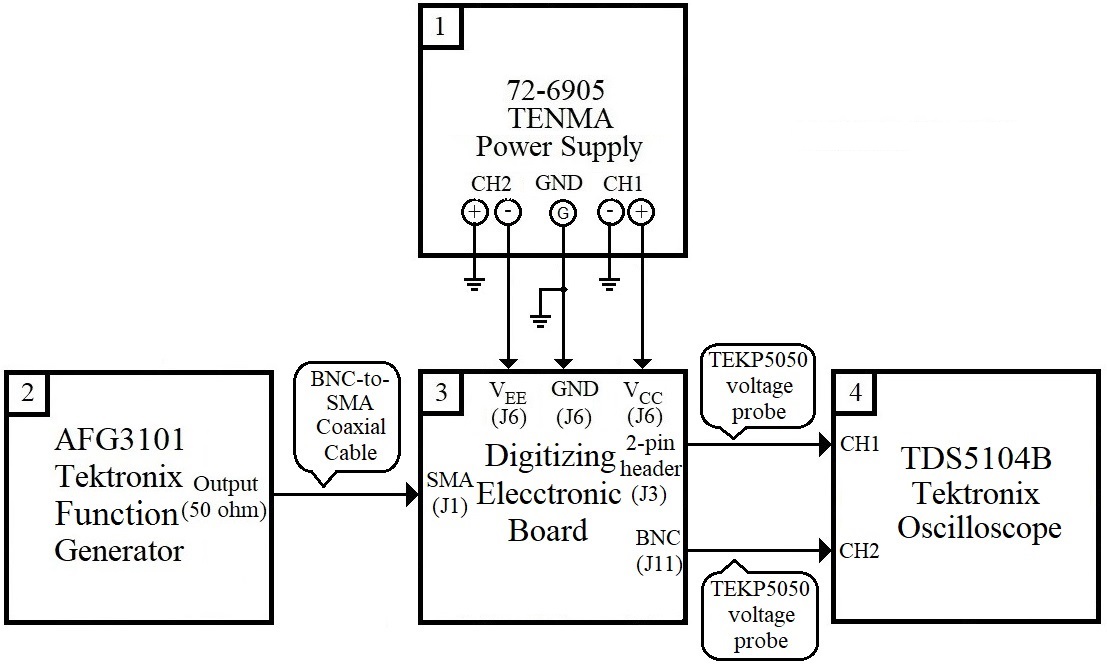}
\caption{Block diagram of the experimental setup to measure digitizing time of the digitizing electronic board.}
\label{FigBloDiaDTime}
\end{figure}

In \textcolor{blue}{Fig. \ref{FigBloDiaDTime}}, we show the block diagram of the digitizing electronic board digitizing time experimental setup.

It contains four blocks as follows:

\begin{enumerate}[1.]
\item BLOCK 1. 72-6905 TENMA power supply.
\item BLOCK 2. AFG3101 Tektronix function generator.
\item BLOCK 3. Digitizing electronic board.
\item BLOCK 4. TDS5104B Tektronix oscilloscope.
\end{enumerate}

BLOCK 1: It represents a 72-6905 TENMA power supply. We used it to generate two output voltages with respect to ground: +5 Vdc (VCC) power supply input and -5 Vdc (VEE) power supply input, to feed the digitizing electronic board.

BLOCK 2: It represents an AFG3101 Tektronix function generator. We used it to supply test analogue signal at the input of the digitizing electronic board.

BLOCK 3: It represents the digitizing electronic board under characterization. It has one analogue input, the SMA connector (J1); and one digital output, BNC connector (J11). See \textcolor{blue}{Fig. \ref{FigSchDiagDigitElecB}}, for schematic diagram of digitizing electronic board; and see \textcolor{blue}{Fig. \ref{FigManufDiagDigitElecB}}, for manufactured digitizing electronic board.

BLOCK 4: It represents a TDS5104B Tektronix oscilloscope. We used it to measure the delay of the digital output signal (CH2) with respect to the analogue input signal (CH1) of the digitizing electronic board.

We used a power cable to connect the digitizing electronic board with the 72-6905 TENMA power supply; power cable features were mentioned in the connection subsection \textcolor{blue}{\ref{EFFconnection}}; the ended stripped wires were connected at the 72-6905 TENMA power supply side; and 3-pin crimp housing receptacle connector, at 3-pin header connector (J6) of the digitizing electronic board side.

We used a BNC-to-SMA coaxial cable to connect the AFG3101 Tektronix function generator with the digitizing electronic board. The BNC goes to the AFG3101 Tektronix functions generator; the SMA, to the SMA connector (J1) of digitizing electronic board.

We used two TEKP5050 Tektronix voltage probes to connect the digitizing electronic board with the TDS5104B Tektronix oscilloscope, as follows:

Connection 1, between 2-pin header connector (J3) \textcolor{blue}{\cite{2-pinTerminalV}} of digitizing electronic board and CH1 of oscilloscope. Probe hook-tip goes to pin number 1; and alligator clip, to SMA metal (J1). The CH1 of the oscilloscope is used to measure the input analogue signal in the digitizing electronic board.

Connection 2, between BNC connector (J11) of the digitizing electronic board and CH2 of oscilloscope. Probe hook-tip goes to contact pin of BNC connector (J11); and alligator clip, to BNC metal (J11). This CH2 of the oscilloscope is used to measure the output digital signal in the digitizing electronic board.

\subsubsection{Measurement processes}
The measurement processes consisted on twelve steps as follows:
\begin{enumerate}[Step 1.]
\item We assembled the experimental setup represented in \textcolor{blue}{Fig. \ref{FigBloDiaDTime}} as a block diagram.

\item We turned on 72-6905 TENMA power supply; adjusted the VCC power supply input at +5 Vdc, and VEE at -5 Vdc.

\item We turned on the measurement equipment (AFG3101 Tektronix function generator, and TDS5104B Tektronix oscilloscope). 50-ohm output connector of the AFG3101 Tektronix function generator was disabled by default.

\item We set up AFG3101 Tektronix function generator, digitizing electronic board, and TDS5104B Tektronix oscilloscope parameters as follows:

\begin{enumerate}[4.1]
\item AFG3101 Tektronix function generator. Waveform: Exponential decay; amplitude: +400 mV\textsubscript{Pk-Pk} ( millivolts peak-to-peak ); load impedance: High-Z; output polarity: Off; voltage offset: 0 Vdc; frequency: 1 MHz; and phase shift: 0.0$^\circ$.
\item Digitizing electronic board. Select trigger polarity (SW3): Positive voltage (shunt connector must be placed between 1 and 2 pins of 3-pin header connector (SW3)); threshold voltage, +100 mVdc; hysteresis option (SW1), on (shunt connector must be placed into hysteresis option (SW1)); hysteresis control variable resistor (R1), 0.9-ohm; internal 50-ohm input impedance option (SW2 and SW4), off (shunt connectors must be released from internal 50-ohm input impedance option (SW2 and SW4)); differential latch input control option (J4), latch mode (first shunt connector must be placed between 1 and 2 pins of differential latch input control option (J4), second shunt connector must be placed between 3 and 4 pins of differential latch input control option (J4)); and install 10-kohm resistor on the 2-pin header connector (J3).
\item TDS5104B Tektronix oscilloscope parameters. CH1 termination: 1-Mohm; CH1 coupling: DC; CH1 invert input signal: Off; CH1 bandwidth: Full; CH1 attenuation: 10X; CH1 vertical scale: 100 mV/DIV; CH1 deskew time: 0.0 s; CH2 termination: 1-Mohm; CH2 coupling: DC; CH2 invert input signal: Off; CH2 bandwidth: Full; CH2 attenuation: 10X; CH2 vertical scale: 1 V/DIV; CH2 deskew time: 0.0 s; horizontal scale: 2 ns/DIV; trigger source: CH1; trigger slope: Rise; trigger mode: Auto; trigger type: Edge; measure 1 menu source and type: CH1, Delay (C1, C2). The Delay (C1, C2) measurement procedure is from C1 to C2 (in time units).
\item The TDS5104B Tektronix oscilloscope front-panel print button was configured as a save button quick storage of screenshots waveform parameters as follows: From the menu bar of the TDS5104B Tektronix oscilloscope, we selected the Save As..., dialog boxes are displayed, we filled the next parameters as follows: Base file name: Defined by us; count: 000; save as type: JPEG; auto-increment file name check box: On. We clicked the Save Button.
\end{enumerate}
\item We enabled the 50-ohm output connector of the AFG3101 Tektronix function generator.
\item We saved at least 20 screenshots of a waveform.
\item We disabled the 50-ohm output connector of the AFG3101 Tektronix function generator; we disconnected the BNC-to-SMA coaxial cable, and two TEKP5050 Tektronix voltage probes from the digitizing electronic board; we turned off 72-6905 TENMA power supply; we released the 3-pin crimp housing receptacle connector of the power supply from 3-pin header connector (J6) of digitizing electronic board; we replaced the digitizing electronic board with the next digitizing electronic board; we connected again the 3-pin crimp housing receptacle connector into the 3-pin header connector (J6) of digitizing electronic board; we turned on 72-6905 TENMA power supply; we connected the BNC-to-SMA coaxial cable into SMA coaxial connector (J1) of digitizing electronic board; we connected two TEKP5050 Tektronix voltage probes: The first connection between 2-pin header connector (J3) of digitizing electronic board and CH1 of oscilloscope, and second connection between BNC connector (J11) of the digitizing electronic board and CH2 of oscilloscope.
\item We repeated the above steps five, six, and seven for each digitizing electronic board. After the fifth digitizing electronic board, we disabled the 50-ohm output connector of the AFG3101 Tektronix function generator; we turned off the AFG3101 Tektronix function generator, TDS5104B Tektronix oscilloscope, 72-6905 TENMA power supply; we disconnected the fifth digitizing electronic board, AFG3101 Tektronix function generator, TDS5104B Tektronix oscilloscope, and 72-6905 TENMA power supply. The process of collecting data for digitizing time of digitizing electronic boards comes to end. The Characterization-4.zip supplementary material folder contains at least 30 screenshots of a waveform of each digitizing electronic board; the delay measurement or Delay (C1, C2) is displayed.

\item We read the Delay (C1, C2) measurement from each screenshot of a waveform, and we wrote into a data text file. This process was not automatic.

\item We repeated the above step nine for each digitizing electronic board screenshot of a waveform. We obtained one data text file for each digitizing electronic board. The Characterization-4.zip supplementary material folder contains five data text files for five digitizing electronics boards.

\item We read five digitizing electronic board data text files and compute the average values and standard deviations of these five data text files, using Statistics02 DTime.vi LabVIEW program, contained in the Characterization-4.zip supplementary material folder.

\item We wrote five statistical error results from Statistics02 DTime.vi LabVIEW program into the script, PlotErrorsDTimeNI-4.c; we run a script, PlotErrorsDTimeNI-4.c, we obtained one plot of the five digitizing electronic boards digitizing time statistical error. The Characterization-4.zip supplementary material folder contains the PlotErrorsDTimeNI-4.c script, and one plot.
\end{enumerate}

\begin{figure}[ht!]
\centering  
\includegraphics[width=468pt]{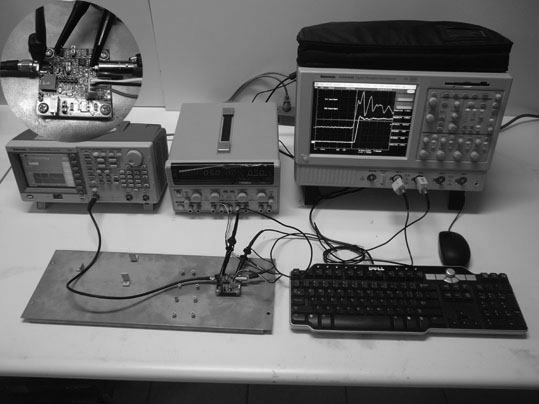}
\caption{Experimental setup to measure the digitizing time of digitizing electronic board.}
\label{FigExpSetDTime}
\end{figure}

In \textcolor{blue}{Fig. \ref{FigExpSetDTime}}, we show the experimental setup to measure the digitizing time of the digitizing electronic board; at the top left side is displayed a figure close up that shows power supply power cable, function generator BNC-to-SMA coaxial cable, and oscilloscope voltage probe CH1 and CH2.

\section{Experimental setup} \label{ExpSetup}
In this section we report on two cases running the experimental setup: First, exposing the experimental setup to cosmic rays. Second, exposing the experimental setup to a controlled LED light source.

\subsection{Exposing the experimental setup to cosmic rays}
In this subsection we report on the connections between the detection material, the connection electronic board, the feeding and reading out electronic board, the digitizing electronic board, the power supplies, and the TDS2022C-EDU Tektronix oscilloscope; these equipment elements conform the experimental setup. We explain the procedure to turn it on. We describe how to configure the digitizing electronic board, the oscilloscope, and the power supplies. We detail the way to run the experimental setup and take data. We inform how to turn off the experimental setup. And we present some results obtained from exposing the detector to cosmic rays.  

With this experimental setup we observe photon-like signals produced in the S12572-100P Hamamatsu photodiode when the detection material is exposed to cosmic rays. We investigated the origin and properties of these signals. Here we use them to test this experimental setup.  

\subsubsection{Connections}
\begin{figure}[ht!]
    \centering
    \includegraphics[width=468pt]{{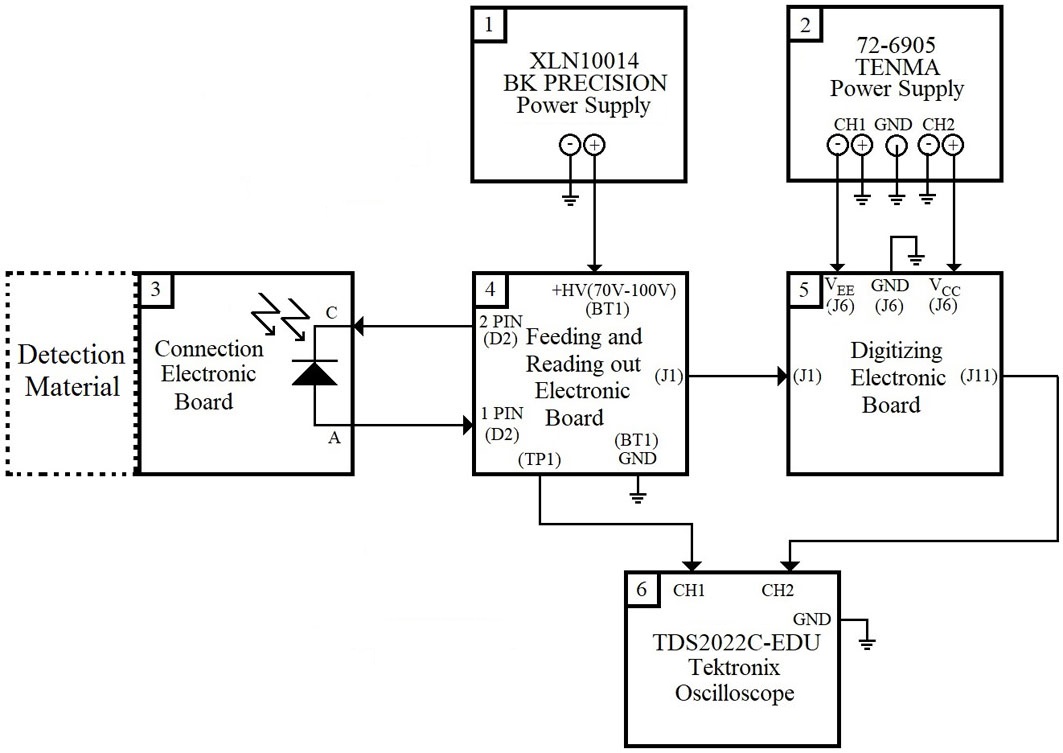}}
    \caption{Block diagram of the complete experimental setup.}
    \label{FigBloDiaExpSet}
\end{figure}

In \textcolor{blue}{Fig. \ref{FigBloDiaExpSet}}, we show the block diagram of the complete experimental setup. 

\textcolor{blue}{Fig. \ref{FigBloDiaExpSet}} contains six blocks as follows:

\begin{enumerate}[1.]
\item BLOCK 1. XLN10014 BK Precision power supply \textcolor{blue}{\cite{UserManual-XLN10014}}.
\item BLOCK 2. 72-6905 TENMA power supply \textcolor{blue}{\cite{Datasheet-72-6905}}.
\item BLOCK 3. Connection electronic board with detection material.
\item BLOCK 4. Feeding and reading out electronic board.
\item BLOCK 5. Digitizing electronic board.
\item BLOCK 6. TDS2022C-EDU Tektronix oscilloscope \textcolor{blue}{\cite{UserManual-TDS2022C-EDU}}.
\end{enumerate}

BLOCK 1: It represents an XLN10014 BK Precision power supply. We used it to feed voltage on the feeding and reading out electronic board, on the 2-pin header connector (BT1) \textcolor{blue}{\cite{2-pinShroud-RA}}, from 0 Vdc to +100 Vdc.

BLOCK 2: It represents a 72-6905 TENMA power supply, which we used to feed the digitizing electronic board with two output voltages: +5 Vdc (VCC) and -5 Vdc (VEE). The prefixes VCC and VEE are defined as power supply inputs in section \textcolor{blue}{\ref{DigElecBoard}}.

BLOCK 3: It represents the connection electronic board. It supports mechanically and electrically the S12572-100P Hamamatsu photodiode, which is attached to the detection material, an 1 in $\times$ 2 in $\times$ 8 in Aluminum bar, see section \textcolor{blue}{\ref{ConnElecBoard}}.

BLOCK 4: It represents the feeding and reading out electronic board. It has two purposes: First, to feed the S12572-100P Hamamatsu photodiode voltage supply using the connection electronic board. Second, to read out the analogue signal from the S12572-100P Hamamatsu photodiode.

BLOCK 5: It represents the digitizing electronic board. It is used to select and fix triggering voltage, and to convert analogue-to-digital signal from feeding and reading out electronic board; the threshold, or triggering, voltage is three times above noise signal.

BLOCK 6: It represents the TDS2022C-EDU Tektronix oscilloscope. We use it to inspect the analogue and digital signals from the feeding and reading out electronic board and digitizing electronic board, respectively.\hfill

We used a power cable to connect the XLN10014 BK Precision power supply with the feeding and reading out electronic board. The characteristics of the cable are as follows: Multiconductor not shielded cable, 24 AWG, two conductors, black jacket (ground or PIN 2 of BT1) and red jacket (+HV(1V-100V) power supply input or PIN 1 of BT1), 300 Vdc voltage rating. The 2-pin crimp housing receptacle connector \textcolor{blue}{\cite{2-pinRectacleCrimp}} goes to the feeding and reading out electronic board; and the solderless type terminal block with screw connector \textcolor{blue}{\cite{XLNTB}} goes to the XLN10014 BK Precision power supply.\hfill

We used a power cable to connect the 72-6905 TENMA power supply with the digitizing electronic board. The characteristics of the power cable are the same as those described in  subsections \textcolor{blue}{\ref{EFFconnection}} and \textcolor{blue}{\ref{DTimeConnection}}. The end stripped wires goes to the 72-6905 TENMA power supply side; and the 3-pin crimp housing receptacle connector \textcolor{blue}{\cite{3-pinRectacleCrimp}}, to the digitizing electronic board.\hfill

\begin{figure}[ht!]
    \centering
    \includegraphics[width=468pt]{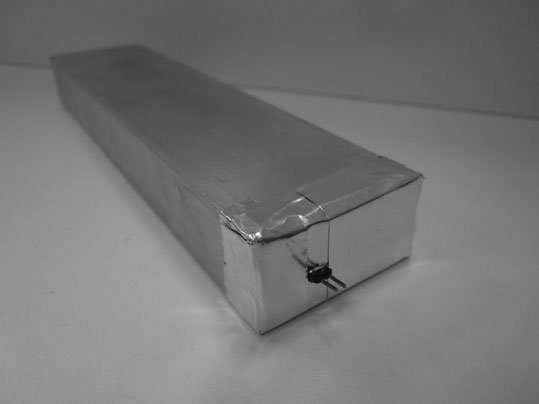}
    \caption{Connection electronic board attached, assembled, and optically isolated with an Aluminum bar.}
    \label{FigAlBar}
\end{figure}

We assembled the Aluminum bar to the connection electronic board as follows: We polished a 1 in $\times$ 2 in end of an 8 in long rectangular Aluminum bar up to a mirror like finishing; we insulated with black electrical tape the soldered anode and cathode pins of the connection electronic board (by the top layer); we applied an additional black electrical tape layer on the top layer of connection electronic board, excluding the optical aperture of the S12572-100P Hamamatsu photodiode; we isolated optically the connection electronic board with one layer of Aluminum tape 3311, except the optical aperture of the S12572-100P Hamamatsu photodiode; we installed the connection electronic board on the above mentioned Aluminum bar 1 in $\times$ 2 in end as follows: The optical aperture of the photodiode touches the polished area of the Aluminum bar, then we fixed the connection electronic board to the Aluminum bar with Aluminum tape 3311, except the 2-pin header connector (anode pin and cathode pin) of the bottom layer. In \textcolor{blue}{Fig. \ref{FigAlBar}}, we show the connection electronic board attached, assembled, and optically isolated with the Aluminum bar.

We used only, without additional board-to-board adapters \textcolor{blue}{\cite{TSW-102-09-T-S-RE}}, a 2-pin header connector of the connection electronic board to connect it to 2-pin receptacle connector (D2) \textcolor{blue}{\cite{2-pinSocketRA}} of the feeding and reading out electronic board.

\begin{figure}[ht!]
    \centering
    \includegraphics[width=468pt]{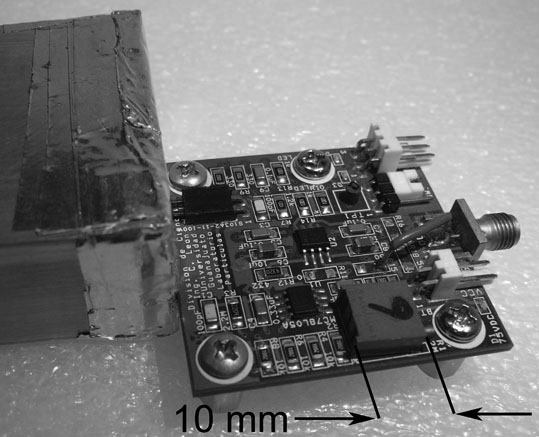}
    \caption{Connection electronic board attached and assembled to the Aluminum bar and connected to the feeding and reading out electronic board.}
    \label{FigAttConElecB}
\end{figure}

In \textcolor{blue}{Fig. \ref{FigAttConElecB}}, we show the connection electronic board attached and assembled to the Aluminum bar and connected to the feeding and reading out electronic board. We used a SMA-to-SMA adapter \textcolor{blue}{\cite{SMApToSMAp}} to connect the feeding and reading out electronic board to the digitizing electronic board, from the SMA connector (J1) of the feeding and reading out electronic board to SMA connector (J1) of the digitizing electronic board.

We used TPP0101 Tektronix voltage probes \textcolor{blue}{\cite{Datasheet-TPP0101}} to connect the feeding and reading out electronic board to the TDS2022C-EDU Tektronix oscilloscope. The probe hook-tip goes to the test point (TP1); and alligator clip, to SMA metal (J1). The BNC goes to channel 1 (CH1) of oscilloscope.

We used a BNC-to-BNC coaxial cable to connect the digitizing electronic board to the TDS2022C-EDU Tektronix oscilloscope. 

\subsubsection{Operations} \label{ExpSetOperation}
The operations to run the experimental setup consist of the following 16 steps:
\begin{enumerate}[Step 1.]
\item We assembled the experimental setup represented in \textcolor{blue}{Fig. \ref{FigBloDiaExpSet}} as a block diagram.
\item We turned on 72-6905 TENMA power supply; adjusted the VCC power supply input to +5 Vdc, and VEE power supply input to -5 Vdc; the output of the 72-6905 TENMA power supply is disabled by default.

\item We set up the digitizing electronic board as follows: Select trigger polarity (SW3), positive voltage (shunt connector must be placed between 1 and 2 pins of 3-pin header connector (SW3)); hysteresis option (SW1), on (shunt connector must be placed into hysteresis option (SW1)); hysteresis control variable resistor (R1), 0.9-ohm; internal 50-ohm input impedance option (SW2 and SW4), off (shunt connectors must be release from internal 50-ohm input impedance option (SW2 and SW4)); differential latch input control option (J4), latch mode (first shunt connector must be placed between 1 and 2 pins of differential latch input control option (J4), second shunt connector must be placed between 3 and 4 pins of differential latch input control option (J4)); and install 10-kohm resistor on the 2-pin header connector (J3).

\item We enable the output of the 72-6905 TENMA power supply by clicking on the ON/OFF output button.

\item We turn on XLN10014 BK Precision power supply; we set up the output at +75 Vdc, and 50 mA; we enable output button. The experimental setup is working.

\item We turn on the TDS2022C-EDU Tektronix oscilloscope; we set up oscilloscope parameters as follows: CH1 coupling, DC; CH1 invert input signal, off; CH1 bandwidth, full; CH1 attenuation, 10X; CH1 vertical scale, 100 mV/DIV; CH2 coupling, DC; CH2 invert input signal, off; CH2 bandwidth, full; CH2 attenuation, 10X; CH2 vertical scale, 2 V/DIV; horizontal scale, 50 ns/DIV; trigger source, CH1; trigger slope, rise; trigger mode, auto; trigger type, edge; measure 1 menu source and type, CH1 -V\textsubscript{Pk-Pk} ( voltage peak-to-peak )-; measure 2 menu source and type, CH1 -rise time-; measure 3 menu source and type, CH1 -fall time-; measure 4 menu source and type, CH2 -V\textsubscript{Pk-Pk} ( voltage peak-to-peak )-; measure 5 menu source and type, CH2 -positive width-.

\item We set up the TDS2022C-EDU Tektronix oscilloscope to save triggers -waveform screenshots- as follows: We insert a USB flash drive into the USB flash drive port for data storage; we push the Save/Recall button to save the oscilloscope screen images, the format selected is JPEG of the save image sub-menu; automatically the file name is generated.

\item We measured noise signal on CH1; the vertical scale of the oscilloscope needed to be adjusted at +2 mV/DIV (only for this step); measurement is not affected if the rising pulse analogue signal goes out of the oscilloscope screen vertical scale; peak-to-peak voltage must be measured before the analogue rising signal appears; after measuring the noise signal on CH1, we set up again the CH1 vertical scale at 100 mV/DIV of TDS2022C-EDU Tektronix oscilloscope (with the same parameter described above in step 6, subsection \textcolor{blue}{\ref{ExpSetOperation}}).

\item We adjusted and fixed the threshold voltage at +60 mVdc (three times bigger above the noise signal, as it was the selection criterion to fix threshold voltage). In subsection \textcolor{blue}{\ref{AdFixThresProcedure}}, we describe in detail the procedure to adjust and fix the threshold voltage.

\item We saved at least 29 screenshots of a waveform by clicking, one time for each screenshot, on the front-panel print button of the oscilloscope; the information was stored in ExperimentalSetup.zip folder, see supplementary material, ExpSet-onHYS-Threshold60mV-PS75Vdc folder.

\item We changed the threshold voltage to +110 mVdc in the digitizing electronic board. We saved at least 26 screenshots of a waveform by clicking one time for each screenshots, on the front-panel print button of the oscilloscope; the information was stored in ExperimentalSetup.zip folder, see supplementary material, ExpSet-onHYS-Threshold110mV-PS75Vdc folder.

\item We changed the threshold voltage to +200 mVdc in the digitizing electronic board. We saved at least 32 screenshots of a waveform by clicking one time for each screenshots, on the front-panel print button of the oscilloscope; the information was stored in ExperimentalSetup.zip folder, see supplementary material, ExpSet-onHYS-Threshold200mV-PS75Vdc folder.

\item We turned off all equipment as follows: First, XLN10014 BK Precision power supply. Second, TDS2022C-EDU Tektronix oscilloscope. Third, 72-6905 TENMA power supply; we disconnected all cables, equipment, and electronic boards.

\item We read, from each screenshot of +60 mVdc threshold voltage, the V\textsubscript{Pk-Pk} of CH1, the rise time of CH1, the fall time of CH1, the V\textsubscript{Pk-Pk} of CH2, the positive width of CH2. We wrote them into a ExpSet-onHYS-Threshold60mV-PS75Vdc.txt data text file. This process was not automatic.

\item We repeated the step fourteen for +110 mVdc threshold voltage, and for +200 mVdc threshold voltage. We obtained one data text file for each threshold voltage; ExpSet-onHYS-Threshold110mV-PS75Vdc.txt for +110 mVdc, and ExpSet-onHYS-Threshold200mV-PS75Vdc.txt for +200 mVdc, respectively. The ExperimentalSetup.zip supplementary material folder contains these three data text files.

\item We read each of the three data text file of threshold voltage -the V\textsubscript{Pk-Pk} of CH1, the rise time of CH1, the fall time of CH1, the V\textsubscript{Pk-Pk} of CH2, the positive width of CH2- and compute their average with ExpSet.c script, contained in ExperimentalSetup.zip folder.
\end{enumerate}

The experimental setup is finished. For collecting data the TDS2022C-EDU Tektronix oscilloscope can be replaced with a data acquisition system.

\begin{figure}[ht!]
    \centering
    \includegraphics[width=468pt]{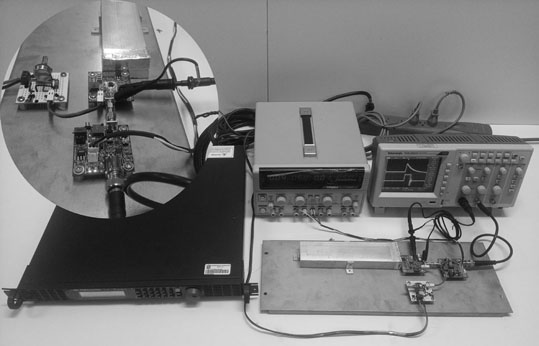}
    \caption{Experimental setup final assembly.}
    \label{FigExpSetAssembly}
\end{figure}

In \textcolor{blue}{Fig. \ref{FigExpSetAssembly}} we show the experimental setup finally assembled; at the top left side corner is displayed a figure close up that shows the power supply’s power cables, and oscilloscope voltage probe CH1 and CH2.

For a video of this experimental setup, please \href{https://laboratoriointernacionaldeparticulaselementales.net/detecci\%C3\%B3n-de-radiaci\%C3\%B3n}{\textcolor{blue}{\underline{click here}}} or visit \textcolor{blue}{\url{https://youtu.be/YcJFaUSOd1U}}

\subsection{Exposing the experimental setup to a controlled LED light source}
In this subsection we report on the experimental setup tests, it consists on exposing the S12572-100P Hamamatsu photodiode to a pulse sequence from a light source (green LED controlled by an AFG3101 Tektronix function generator) in two different experimental setups: Complete experimental setup (which is formed by connection electronic board, feeding and reading out electronic board, and digitizing electronic board); partial experimental setup (which is formed by connection electronic board, and feeding and reading out electronic board). The pulse sequence is produced by the pulsating light source from the experimental setup tests, a digital output was measured in an environment isolated from electromagnetic waves (wooden box painted black inside, coated with Aluminum tape outside, and hermetically sealed from electromagnetic radiation). The LED light source was excited with pulse waveform (amplitude: 4 Vpp ( voltage peak-to-peak ); offset: 0 Vdc) and exponential decay waveform (amplitude: 4 Vpp ( voltage peak-to-peak ); offset: 0 Vdc; duty cycle: 1.5\%). The feeding voltage of +80 Vdc was used on the feeding and reading out electronic board. The feeding voltage of +5 Vdc (VCC) power supply input, -5 Vdc (VEE) power supply input, and threshold voltage of +20 mVdc was using on the digitizing electronic board. The same parameters were used in all the experimental setup tests.

The experimental setup tests were evaluated for two different waveform responses as follows:

\begin{enumerate}[1.]
\item Exponential decay waveform response to the LED light source.

Two preliminary results were obtained as follows:
\begin{enumerate}
\item Partial experimental setup.

The S12572-100P Hamamatsu photodiode in the connection electronic board and feeding and reading out electronic board worked very well for frequencies between 1 Hz up to 1 MHz.

\item Complete experimental setup.

The S12572-100P Hamamatsu photodiode in the connection electronic board, feeding and reading out electronic board, and digitizing electronic board worked for frequencies between 1 Hz up to 10 kHz.
\end{enumerate}
\item Pulse waveform response to the LED light source.

Two preliminary results were obtained as follows:
\begin{enumerate}
\item Partial experimental setup.

The S12572-100P Hamamatsu photodiode in the connection electronic board and feeding and reading out electronic board worked very well for frequencies between 1 Hz up to 2 MHz.

\item Complete experimental setup.

The S12572-100P Hamamatsu photodiode in the connection electronic board, feeding and reading out electronic board, and digitizing electronic board worked for frequencies between 1 Hz up to 100 kHz.
\end{enumerate}
\end{enumerate}

The offset voltage of -1 Vdc at frequencies close to 100 kHz (pulse waveform) and 10 kHz (exponential decay waveform) between the connection of feeding and reading out electronic board and digitizing electronic board were measured; by increasing the LED light source frequency, an offset voltage different from 0 Vdc was observed. The 10-kohm resistor (J3) in conjunction with the comparator circuit (U1) in the digitizing electronic board has a limit of operation when connected to STAGE 1 circuit of the feeding and reading out electronic board; an improvement in the electronic coupling between the two electronic boards will be required.

\section{Results}
In this section, we present the results of the front-end electronic boards characterization -attenuation factor, phase shift, transit time of the feeding and reading out electronic board, digitizing efficiency error, digitizing efficiency, digitizing time of digitizing electronic board - and experimental setup -three screenshot samples corresponding to +60 mVdc, +110  mVdc, and +200 mVdc of threshold voltage, using a feeding voltage of +75 Vdc; and the average values from all the screenshots of V\textsubscript{Pk-Pk} of CH1, rise time of CH1, fall time of CH1, V\textsubscript{Pk-Pk} of CH2, positive width of CH2-.

\subsection {Results from front-end electronic boards' characterization}
We show six characterization results as follows: 
\begin{enumerate}[1.]
\item Feeding and reading out electronic board attenuation factor.
\item Feeding and reading out electronic board phase shift.
\item Feeding and reading out electronic board transit time.
\item Digitizing-electronic-board digitizing efficiency error.
\item Digitizing-electronic-board digitizing efficiency.
\item Digitizing-electronic board digitizing time.
\end{enumerate}

\subsubsection{Results from feeding and reading out electronic board attenuation factor}
In this subsection, we show the results of the attenuation factor from feeding and reading out electronic board number 1, as a function of the injected signal frequency.

We displayed the data from the data text file -2018-04-20 1CH T1.txt, from Characterization-1.zip folder- in three graphs: Attenuation factor as function of the frequency between 100 Hz to 10 kHz, attenuation factor as function of the frequency between 10 kHz to 1 MHz, and attenuation factor as function of the frequency between 100 Hz to 1 MHz; all of them include the linear fit parameters and the $\chi^{2}$ per dof (the number of degrees of freedom), p0 (the Y intercept of the first order polynomial fit and, physically, the average attenuation factor of the feeding and reading out electronic board number 1), and p1 (the slope). The p0's are very similar, in the average it is 0.78; the p1's are near to zero; the best fit is linear in the three cases. 

\begin{figure}[ht!]
    \centering
    \includegraphics[width=350pt]{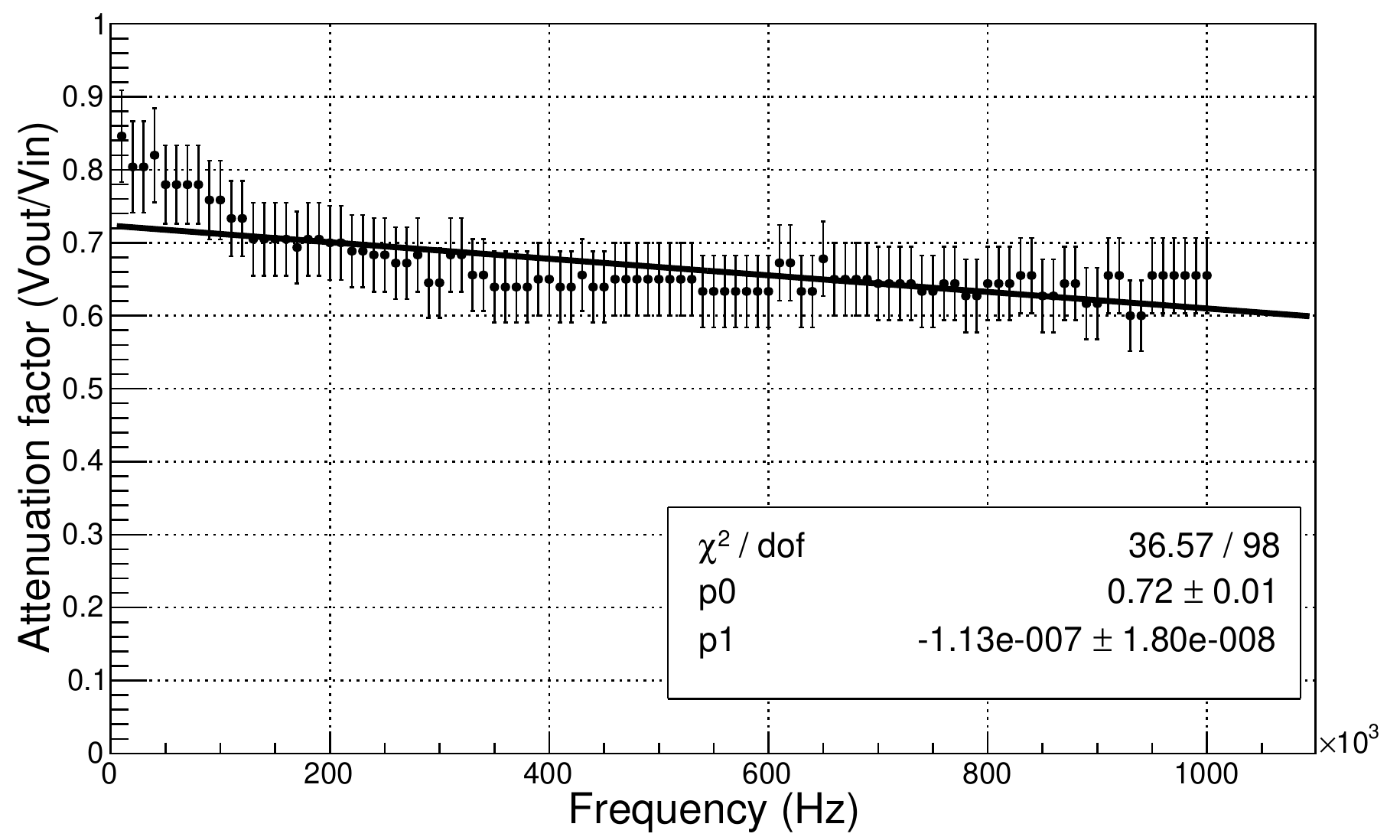}
    \caption{Attenuation factor as a function of the frequency between 10 kHz to 1 MHz from feeding and reading out electronic board number 1; superimposed linear fit.}
    \label{FigAFremaining}
\end{figure}

\begin{figure}[ht!]
    \centering
    \includegraphics[width=350pt]{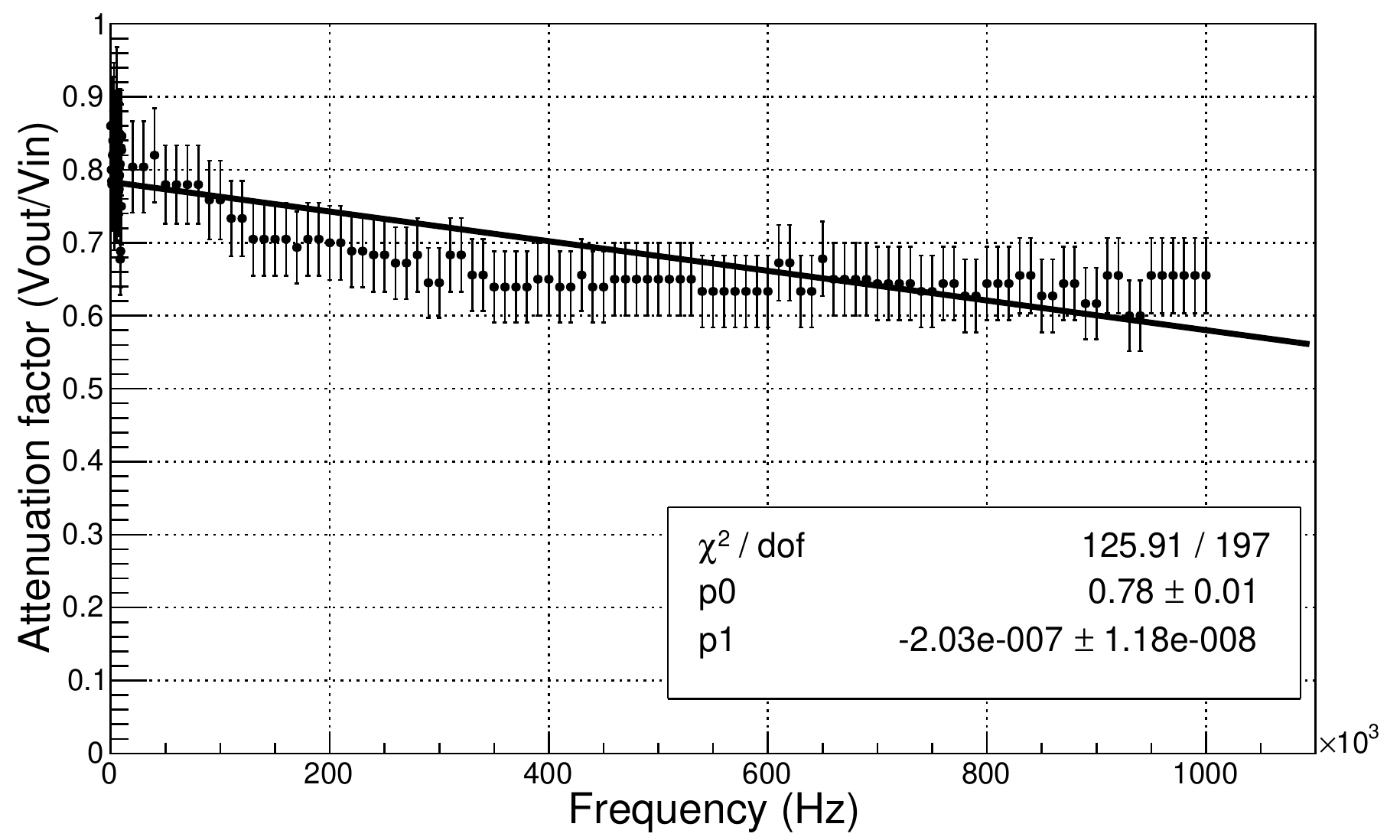}
    \caption{Attenuation factor as a function of the frequency between 100 Hz to 1 MHz from feeding and reading out electronic board number 1; superimposed linear fit.}
    \label{FigAF199M}
\end{figure}

In \textcolor{blue}{Fig. \ref{FigAF100Mfirst}}, \textcolor{blue}{Fig. \ref{FigAFremaining}}, and \textcolor{blue}{Fig. \ref{FigAF199M}}, we show the results of the attenuation factor as a function of frequency; from 100 Hz to 10 kHz (the first 100 measurements), 10 kHz to 1 MHz (the remaining measurements), and 100 Hz to 1 MHz frequencies (199 measurements), respectively.

The fit parameters are as follows:

\begin{figure}[ht!]
    \centering
    \includegraphics[width=350pt]{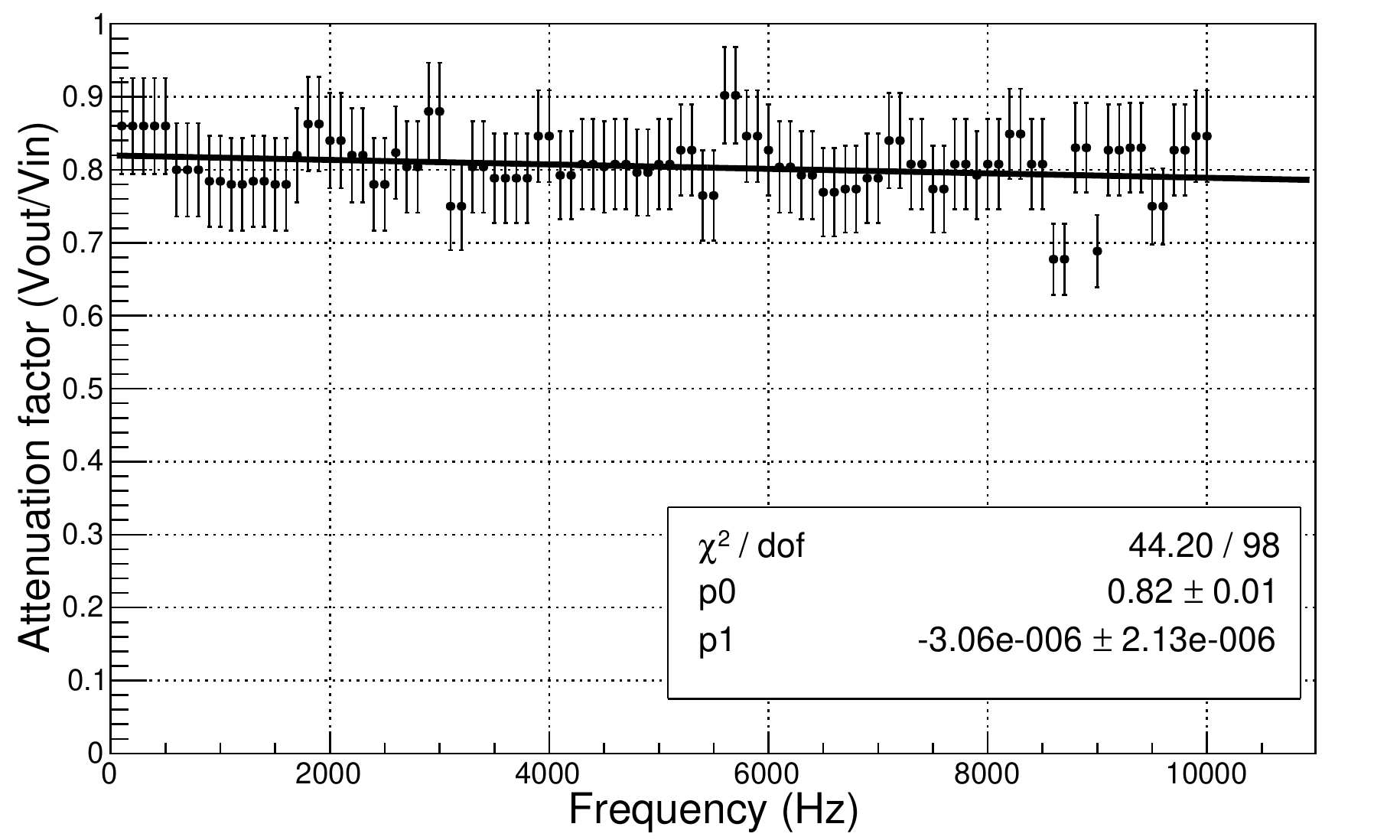}
    \caption{Attenuation factor as a function of the frequency between 100 Hz to 10 kHz from feeding and reading out electronic board number 1; superimposed linear fit.}
    \label{FigAF100Mfirst}
\end{figure}

In the range of frequencies from 100 Hz to 10 kHz: $\chi^{2}$ is 44.20, dof is 98, p0 is 0.82$\pm$0.01, and p1 is (-3.06$\pm$2.13) $\times 10^{-6}$. 

In the range of frequencies from 10 kHz to 1 MHz: $\chi^{2}$ is 36.57, dof is 98, p0 is 0.72 $\pm$0.01, and p1 is (-11.30$\pm$1.80) $\times 10^{-8}$. 

In the range of frequencies from 100 Hz to 1 MHz, are as follows: $\chi^{2}$ is 125.91, dof is 197, p0 is 0.78 $\pm$0.01, and p1 is (-20.30$\pm$1.18) $\times 10^{-8}$. 

We repeated the characterization procedure for the attenuation factor as a function of frequency for feeding and reading out electronic boards number 2 to 10; we plot in 27 graphs the results; nine graphs from 100 Hz to 10 kHz (the first 100 measurements), nine graphs from 10 kHz to 1 MHz (the remaining measurements), and nine graphs from 100 Hz to 1 MHz (199 measurements).

In \textcolor{blue}{Table \ref{table:AF-100MeasurementsF}}, \textcolor{blue}{Table \ref{table:AF-rMeasurements}} and \textcolor{blue}{Table \ref{table:AF-199Measurements}}, we show the results of the first-order polynomial fit, of the attenuation factor from ten feeding and reading out electronic boards, in frequencies from 100 Hz to 10kHz (the first 100 measurements), in frequencies from 10 kHz to 1 MHz (the remaining measurements), and frequencies from 100 Hz to 1 MHz (199 measurements), respectively.

\begin{table}[ht!]
\caption{Attenuation factor as a function of frequencies from 100 Hz to 10 kHz (the first 100 measurements) for ten feeding and reading out electronic boards.}
\centering
\begin{tabular}{rrccrc}
\hline\hline                      
\begin{tabular}[c]{@{}c@{}}Board\\number\end{tabular} &
\begin{tabular}[c]{@{}c@{}}$\chi^{2}$\;\;\end{tabular} & 
\begin{tabular}[c]{@{}c@{}}dof\end{tabular} &
\begin{tabular}[c]{@{}c@{}}p0\end{tabular} &
\begin{tabular}[c]{@{}c@{}}p1\;\;\;\;\;\;\\ ($\times 10^{-9}$)\;\;\;\;\;\;\end{tabular} & \\
\hline	  
1\;\;\;\;\; & 44.20 & 98 & 0.82$\pm$0.01 & -3060.00$\pm$2130.00\\
2\;\;\;\;\; & 59.11 & 98 & 0.80$\pm$0.01 & -2320.00$\pm$2100.00\\
3\;\;\;\;\; & 65.57 & 98 & 0.81$\pm$0.01 & -2650.00$\pm$2120.00\\
4\;\;\;\;\; & 59.75 & 98 & 0.82$\pm$0.01 & +1930.00$\pm$2180.00\\
5\;\;\;\;\; & 115.99 & 98 & 0.79$\pm$0.01 & -2640.00$\pm$2060.00\\
6\;\;\;\;\; & 59.74 & 98 & 0.78$\pm$0.01 & -8.15$\pm$2100.00\\
7\;\;\;\;\; & 75.98 & 98 & 0.78$\pm$0.01 & -870.00$\pm$2060.00\\
8\;\;\;\;\; & 66.61 & 98 & 0.78$\pm$0.01 & +509.00$\pm$2110.00\\
9\;\;\;\;\; & 88.25 & 98 & 0.78$\pm$0.01 & -1180.00$\pm$2040.00\\
10\;\;\;\;\; & 55.54 & 98 & 0.80$\pm$0.01 & -1830.00$\pm$2070.00\\
\hline
\hline                     
\end{tabular}
\label{table:AF-100MeasurementsF}
\end{table}

\begin{table}[ht!]
\caption{Attenuation factor as a function of frequencies from 10 kHz to 1 MHz (the remaining measurements) for ten feeding and reading out electronic boards.}
\centering
\begin{tabular}{rcccrc}
\hline\hline                      
\begin{tabular}[c]{@{}c@{}}Board\\ number\end{tabular} & 
\begin{tabular}[c]{@{}c@{}}$\chi^{2}$\end{tabular} & 
\begin{tabular}[c]{@{}c@{}}dof\end{tabular} &
\begin{tabular}[c]{@{}c@{}}p0\end{tabular} &
\begin{tabular}[c]{@{}c@{}}p1\;\;\\($\times 10^{-8}$)\;\;\end{tabular} & \\
\hline	  
1\;\;\;\;\; & 36.57 & 98 & 0.72$\pm$0.01 & -11.30$\pm$1.80\\
2\;\;\;\;\; & 37.83 & 98 & 0.73$\pm$0.01 & -12.50$\pm$1.79\\
3\;\;\;\;\; & 46.56 & 98 & 0.72$\pm$0.01 & -11.20$\pm$1.79\\
4\;\;\;\;\; & 27.52 & 98 & 0.75$\pm$0.01 & -8.59$\pm$1.82\\
5\;\;\;\;\; & 41.09 & 98 & 0.73$\pm$0.01 & -11.90$\pm$1.80\\
6\;\;\;\;\; & 43.67 & 98 & 0.73$\pm$0.01 & -11.40$\pm$1.82\\
7\;\;\;\;\; & 41.42 & 98 & 0.73$\pm$0.01 & -12.80$\pm$1.81\\
8\;\;\;\;\; & 58.04 & 98 & 0.72$\pm$0.01 & -10.00$\pm$1.81\\
9\;\;\;\;\; & 37.14 & 98 & 0.72$\pm$0.01 & -11.30$\pm$1.79\\
10\;\;\;\;\; & 37.33 & 98 & 0.73$\pm$0.01 & -11.50$\pm$1.80\\
\hline
\hline                     
\end{tabular}
\label{table:AF-rMeasurements}
\end{table}

\begin{table}[ht!]
\caption{Attenuation factor as a function of frequencies from 100 Hz to 1 MHz (199 measurements) for ten feeding and reading out electronic boards.}
\centering
\begin{tabular}{rccccc}
\hline\hline                      
\begin{tabular}[c]{@{}c@{}}Board\\ number\end{tabular} & 
\begin{tabular}[c]{@{}c@{}}$\chi^{2}$\end{tabular} & 
\begin{tabular}[c]{@{}c@{}}dof\end{tabular} &
\begin{tabular}[c]{@{}c@{}}p0\end{tabular} &
\begin{tabular}[c]{@{}c@{}}p1\\($\times 10^{-8}$)\end{tabular} & \\
\hline	  
1\;\;\;\;\; & 125.91 & 197 & 0.78$\pm$0.01 & -20.30$\pm$1.18\\
2\;\;\;\;\; & 127.99 & 197 & 0.78$\pm$0.01 & -19.90$\pm$1.17\\
3\;\;\;\;\; & 152.20 & 197 & 0.77$\pm$0.01 & -19.60$\pm$1.17\\
4\;\;\;\;\; & 138.30 & 197 & 0.81$\pm$0.01 & -18.60$\pm$1.20\\
5\;\;\;\;\; & 170.26 & 197 & 0.76$\pm$0.01 & -17.80$\pm$1.17\\
6\;\;\;\;\; & 126.96 & 197 & 0.77$\pm$0.01 & -18.20$\pm$1.18\\
7\;\;\;\;\; & 130.71 & 197 & 0.76$\pm$0.01 & -17.80$\pm$1.17\\
8\;\;\;\;\; & 149.85 & 197 & 0.77$\pm$0.01 & -17.00$\pm$1.18\\
9\;\;\;\;\; & 147.19 & 197 & 0.76$\pm$0.01 & -17.70$\pm$1.16\\
10\;\;\;\;\; & 120.13 & 197 & 0.78$\pm$0.01 & -18.60$\pm$1.17\\
\hline
\hline                     
\end{tabular}
\label{table:AF-199Measurements}
\end{table}

The complete information, ten data text files, three scripts, and 30 graphics with the first-order polynomial of the attenuation factor included, from ten feedings and reading out electronic boards can be found inside the Characterization-1.zip supplementary material folder.

\subsubsection{Results from feeding and reading out electronic board phase shift}
In this subsection, we show the results of the phase shift from feeding and reading out electronic board number 1, as a function of the injected signal frequency; we showed a graph of the data text file -2018-04-20 1CH T01.txt- phase shift in the range of 100 Hz to 1 MHz (199 measurements).

\begin{figure}[ht!]
    \centering
    \includegraphics[width=400pt]{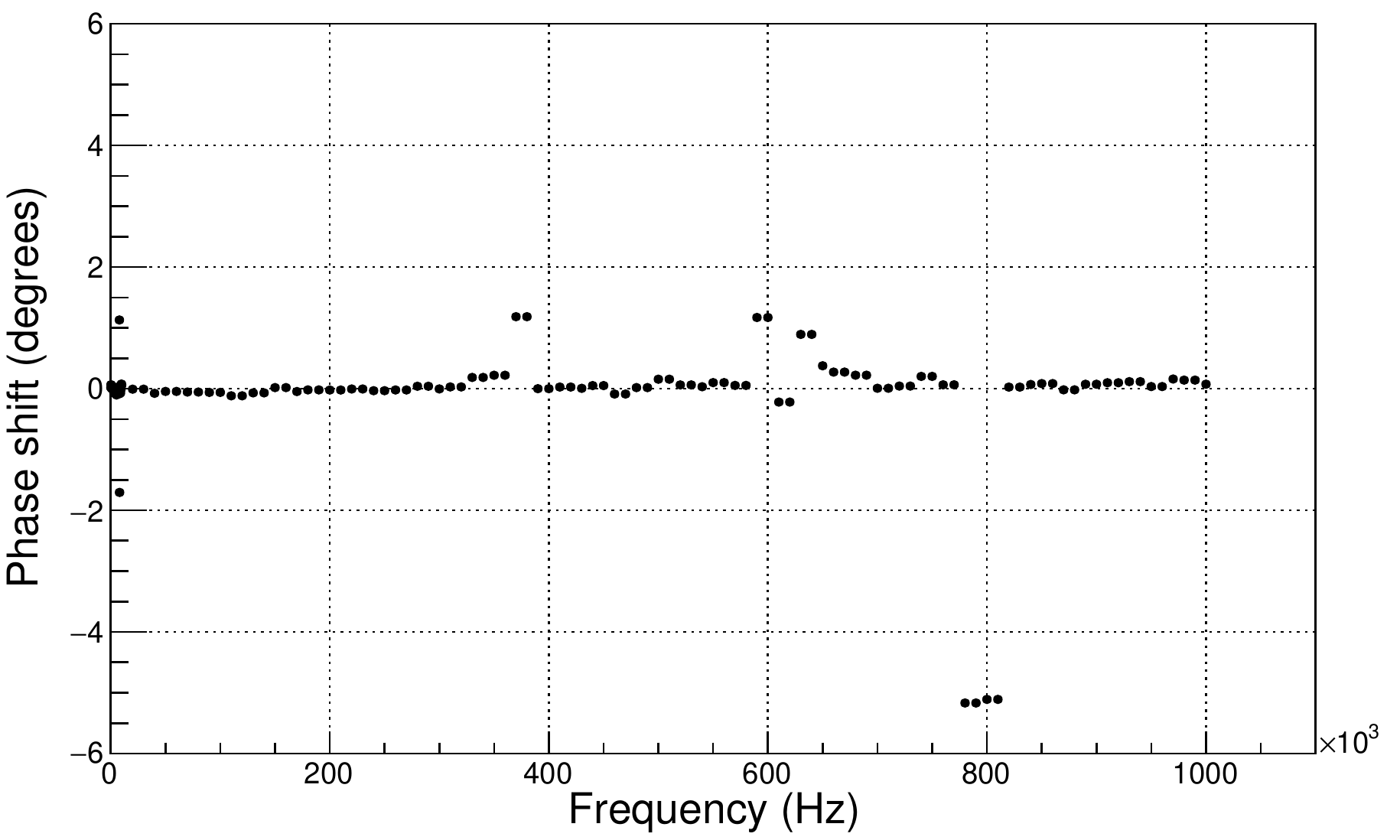}
    \caption{Phase shift measurements vs frequency for 100 Hz to 1 MHz from feeding and reading out electronic board number 1; linear scale.}
    \label{FigPS199mLinS}
\end{figure}

\begin{figure}[ht!]
    \centering
    \includegraphics[width=400pt]{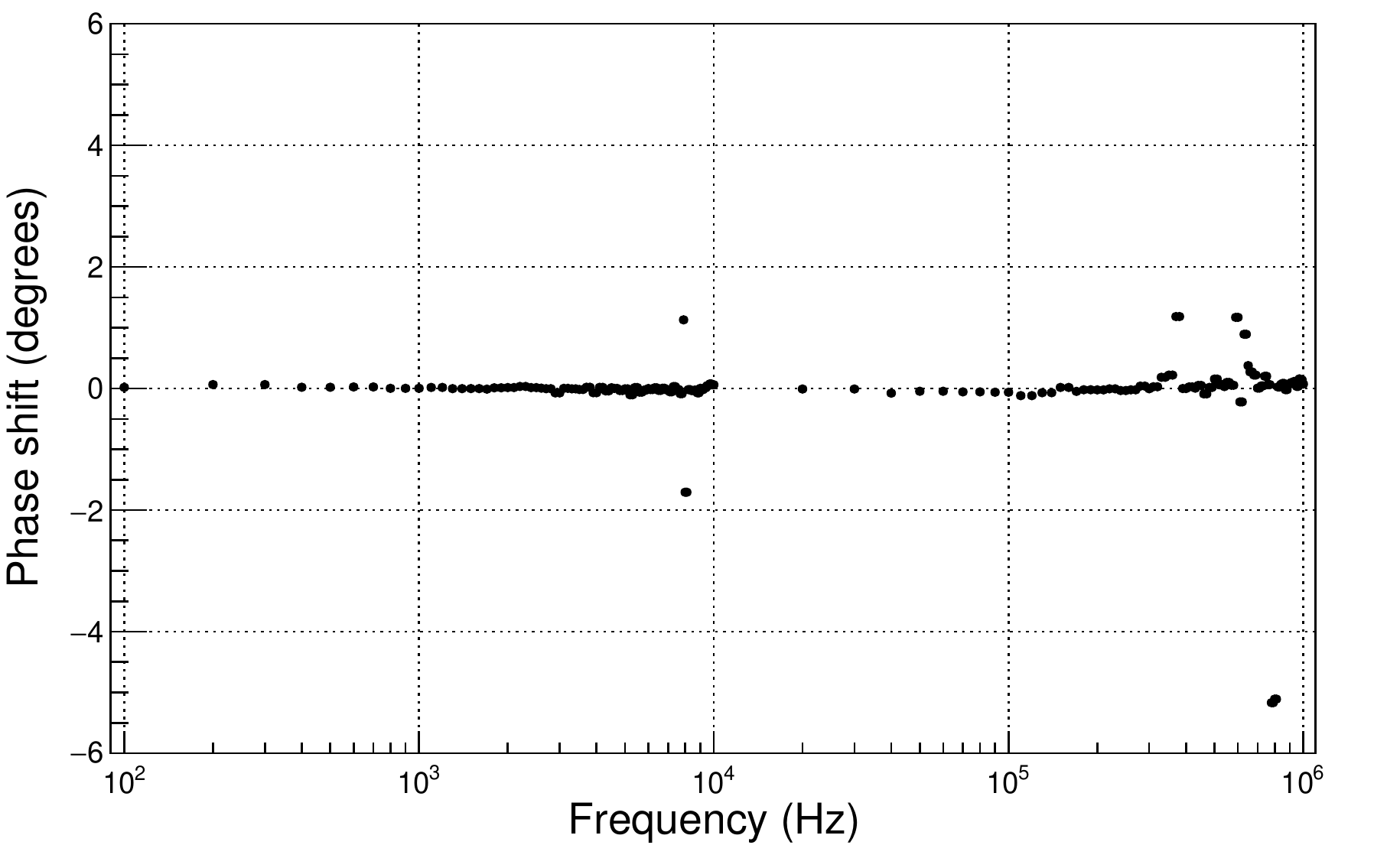}
    \caption{Phase shift measurements vs frequency for 100 Hz to 1 MHz from feeding and reading out electronic board number 1; logarithmic scale.}
    \label{FigPS199mLogS}
\end{figure}

In \textcolor{blue}{Fig. \ref{FigPS199mLinS}}, we show the results of the phase shift as a function of frequency for 100 Hz to 1 MHz (199 measurements) from feeding and reading out electronic board number 1; the scale is linear; at low frequencies, 100 Hz to 10 kHz, there are two values for the phase shift less than two degrees, not clearly distinguished; at higher frequencies, 10 kHz to 1 MHz, there are four values for the phase shift higher than two degrees and less than six degrees, clearly distinguished. In \textcolor{blue}{Fig. \ref{FigPS199mLogS}}, we show the results in logarithmic scale, the phase shifts at low frequencies are clearly distinguished.

We repeated the characterization procedure of the phase shift as a function frequency for the feeding and reading out electronic boards number 2 to 10; we got the results of the attenuation factor as a function of frequency from 100 Hz to 1 MHz (199 measurements, for each electronic board); we display the results in 18 graphs, two for each feeding and reading out electronic board; the 20 graphs (ten linear scale, and ten logarithmic scale for each of the ten feeding and reading out electronic boards) can be found inside the Characterization-1.zip supplementary material folder.

In \textcolor{blue}{Table \ref{table:PhaseShift}}, we show the phase shift results for ten feeding and reading out electronic boards; phase shift reported measurements are equal or greater than 2 degrees. 

\begin{table}[!hbp]
\caption{Phase shift for ten feeding and reading out electronic boards; phase shift measurements reported are equal or greater than 2 degrees.}
\centering
\begin{tabular}{rrrcrrcrrc}
\hline\hline
\begin{tabular}[c]{@{}c@{}}Board\\ number\end{tabular} & 
\begin{tabular}[c]{@{}c@{}}Frequency\\ (kHz)\end{tabular} & 
\begin{tabular}[c]{@{}c@{}}Phase\\ shift \\ (Deg.)\end{tabular} &
\begin{tabular}[c]{@{}c@{}}Board\\ number\end{tabular} & 
\begin{tabular}[c]{@{}c@{}}Frequency\\ (kHz)\end{tabular} &
\begin{tabular}[c]{@{}c@{}}Phase\\ shift \\ (Deg.)\end{tabular} &
\begin{tabular}[c]{@{}c@{}}Board\\ number\end{tabular} &
\begin{tabular}[c]{@{}c@{}}Frequency\\ (kHz)\end{tabular} &
\begin{tabular}[c]{@{}c@{}}Phase\\ shift\\ (Deg.)\end{tabular} & \\
\hline
\hline
  & 780.0\;\;\;\; & -5.17\; & & 4.0\;\;\;\; & -3.91\; & & 7.9\;\;\;\; & +5.06\;\\
  & 790.0\;\;\;\; & -5.17\; & & 4.1\;\;\;\; & -3.91\; & & 8.0\;\;\;\; & +5.06\;\\
1\;\;\;\; & 800.0\;\;\;\; & -5.11\; & 2 & 790.0\;\;\;\; & -5.33\; & 3 & 390.0\;\;\;\; & -36.22\;\\
  & 810.0\;\;\;\; & -5.11\; & & 800.0\;\;\;\; & -5.33\; & & 400.0\;\;\;\; & -36.22\;\\ 
  &       &       &   &        &       &   & 800.0\;\;\;\; & -5.99\;\\
  &       &       &   &        &       &   & 810.0\;\;\;\; & -5.99\;\\
 \hline
  & 4.0\;\;\;\; & +6.13\; &   &   7.9\;\;\;\; & +3.94\; &   &   6.0\;\;\;\; & +3.39\;\\
  & 4.1\;\;\;\; & +6.13\; &   &   8.0\;\;\;\; & +3.94\; &   &   8.0\;\;\;\; & +3.68\;\\
  & 8.0\;\;\;\; & +3.42\; &   & 400.0\;\;\;\; & -7.59\; &   &   8.1\;\;\;\;  & +3.68\;\\
4\;\;\;\; & 8.1\;\;\;\; & +3.42\; & 5 & 410.0\;\;\;\; & -7.59\; & 6 & 400.0\;\;\;\; & +4.04\;\\ 
  & 660.0\;\;\;\; & +2.11\; &   & 790.0\;\;\;\; & -4.49\; &   & 410.0\;\;\;\; & +4.04\;\\
  & 670.0\;\;\;\; & +2.11\; &   & 800.0\;\;\;\; & -4.49\; &   & 790.0\;\;\;\; & -4.49\;\\
  & 800.0\;\;\;\; & -4.55\; &   &       &       &   & 800.0\;\;\;\; & -4.49\;\\
  & 810.0\;\;\;\; & -4.55\; &   &       &       &   &       & \\
 \hline
  &   4.0\;\;\;\; & +3.92\; &   &   6.0\;\;\;\; & -2.55\; &   &   4.0\;\;\;\; & +3.90\;\\
  &   4.1\;\;\;\; & +3.92\; &   &   6.1\;\;\;\; & -2.55\; &   &   4.1\;\;\;\; & +3.90\;\\
  & 790.0\;\;\;\; & -3.93\; &   & 400.0\;\;\;\; & -7.31\; &   &   8.0\;\;\;\; & +5.39\;\\
7\;\;\;\; & 800.0\;\;\;\; & -3.93\; & 8 & 410.0\;\;\;\; & -7.31\; & 9 &   8.1\;\;\;\; & +5.39\;\\ 
  &       &       &   & 790.0\;\;\;\; & -5.99\; &   & 400.0\;\;\;\; & +4.65\;\\
  &       &       &   & 800.0\;\;\;\; & -5.99\; &   & 410.0\;\;\;\; & +4.65\;\\
  &       &       &   &       &       &   & 790.0\;\;\;\; & -4.03\;\\
  &       &       &   &       &       &   & 800.0\;\;\;\; & -4.03\;\\
 \hline
   &   8.0\;\;\;\; & +2.84\; & & & & & &\\
   & 780.0\;\;\;\; & -5.26\; & & & & & &\\
10\;\;\;\; & 790.0\;\;\;\; & -5.26\; & & & & & &\\
   & 800.0\;\;\;\; & -3.78\; & & & & & &\\ 
   & 810.0\;\;\;\; & -3.78\; & & & & & &\\
\hline
\hline   	
\end{tabular}
\label{table:PhaseShift}
\end{table}

From the results of \textcolor{blue}{Table \ref{table:PhaseShift}}, we observe that the phase shift is not generated at all frequencies, but were randomly produced at some frequencies, with very small values, occasionally with high values. These are some observed values: A phase shift between -3.91$^\circ$ and +6.13$^\circ$ at the input analogue signal frequency between 4 Hz to 8.1 Hz, from the feeding and reading out electronic boards numbers 2 to 10. A phase shift between -7.59$^\circ$ and +4.65$^\circ$ at the injected analogue signal frequency near 400 kHz, from the feeding and reading out electronic boards numbers 5, 6, 8, and 9. A phase shift of -36.22$^\circ$ at the injected analogue signal frequency 390 kHz and 400 kHz, from the feeding and reading out electronic board number 3. A phase shift between -5.99$^\circ$ and -3.78$^\circ$ at the injected analogue signal frequency near 800 kHz, from feeding and reading out electronic boards numbers 1 to 10. 

\subsubsection{Results from feeding and reading out electronic board transit time}
We show the results from the subsection \textcolor{blue}{\ref{TTimeC}} of the time an electrical signal takes to travel through the electronic channel of each feeding and reading out electronic board.

\begin{figure}[ht!]
    \centering
    \includegraphics[width=468pt]{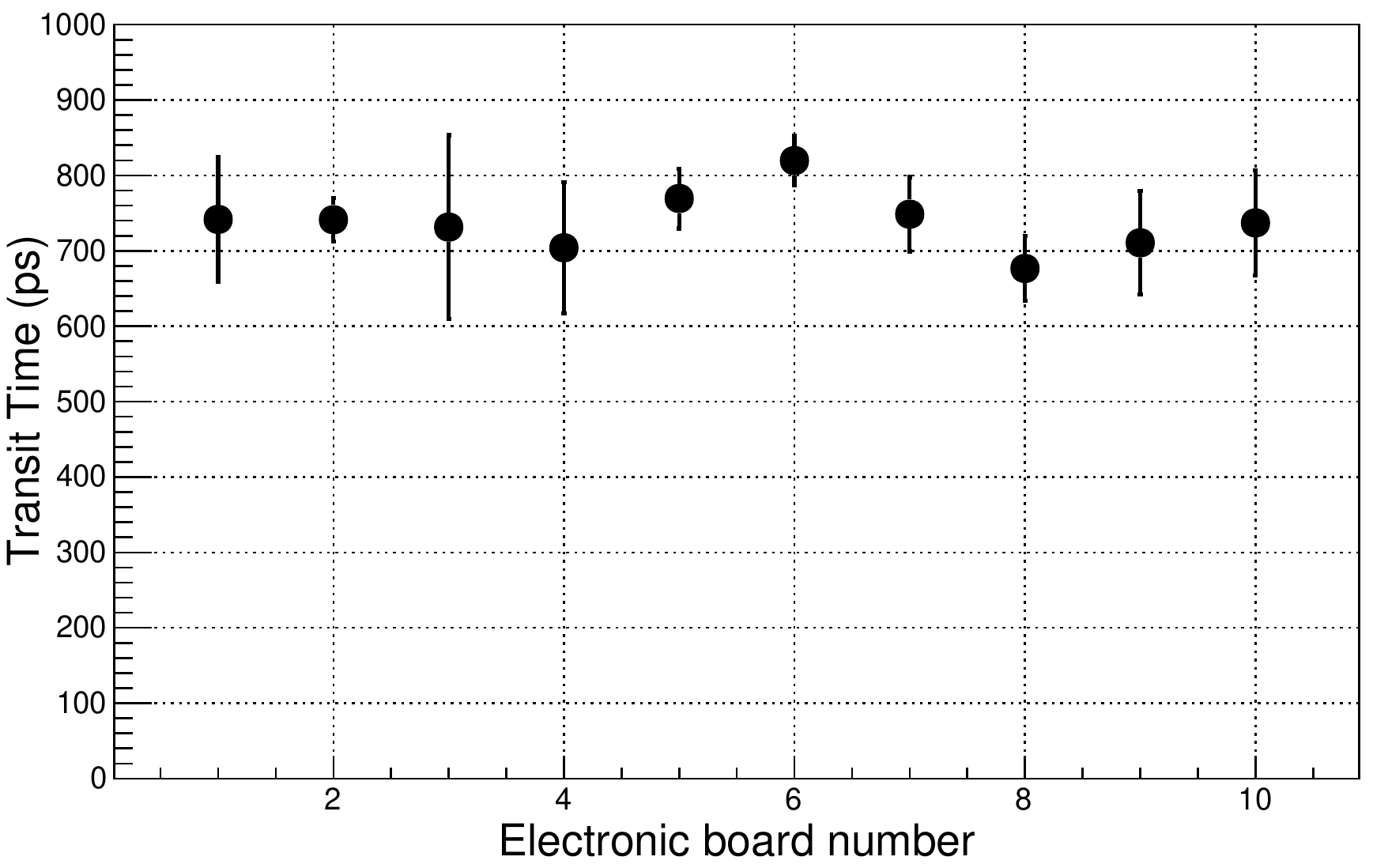}
    \caption{Transit time vs electronic board number, from \textcolor{blue}{Table \ref{table:TTime}}, statistical errors only.}
    \label{FigTTime}
\end{figure}

In \textcolor{blue}{Table \ref{table:TTime}} and \textcolor{blue}{Fig. \ref{FigTTime}}, we show the transit time and its error for ten feeding and reading out electronic boards. The overall transit time is approximately 738.08$\pm$62.38 ps; it is an average of 10 feeding and reading out electronic boards transit time from \textcolor{blue}{Table \ref{table:TTime}}.

\begin{table}[!hbp]
\caption{Transit time for ten feeding and reading out electronic boards.}
\centering
\begin{tabular}{r r c}
\hline\hline                      
\begin{tabular}[c]{@{}c@{}}Board\\ number\end{tabular} & 
\begin{tabular}[c]{@{}c@{}}Transit time\\ (ps)\end{tabular} & \\
\hline          
1\;\;\;\;\; & +741.72$\pm$\;\;82.80 \\
2\;\;\;\;\; & +741.27$\pm$\;\;29.07 \\
3\;\;\;\;\; & +731.52$\pm$121.87 \\
4\;\;\;\;\; & +704.07$\pm$\;\;86.75 \\
5\;\;\;\;\; & +769.46$\pm$\;\;39.80 \\
6\;\;\;\;\; & +819.75$\pm$\;\;33.06 \\
7\;\;\;\;\; & +748.60$\pm$\;\;49.21 \\
8\;\;\;\;\; & +676.65$\pm$\;\;42.79 \\
9\;\;\;\;\; & +710.68$\pm$\;\;68.59 \\
10\;\;\;\;\; & +737.05$\pm$\;\;69.82 \\
\hline 
\hline                     
\end{tabular}
\label{table:TTime}
\end{table}

\subsubsection{Results from digitizing-electronic-board digitizing efficiency error}
In this subsection, we show the results of digitizing efficiency error ($eff$) Eq. \textcolor{blue}{\eqref{eq:err}} from five digitizing electronic boards and AFG3101 Tektronix function generator. When the pulses per second measured at the output of the digitizing electronic board are equal to the input pulses per second, the $eff$ is zero. We observe that the $eff$ is a function of input frequency and the digitizing electronic boards. These are some observed results: At frequencies between 1 Hz to 10 Hz, the $eff$ is zero; at frequencies between 100 Hz to 1 MHz, the $eff$ is increased; the digitizing electronic board number 2 showed the lowest $eff$ and very close to AFG3101 Tektronix function generator $eff$; the digitizing electronic board number 1 has greater $eff$.

\begin{figure}[ht!]
    \centering
    \includegraphics[width=400pt]{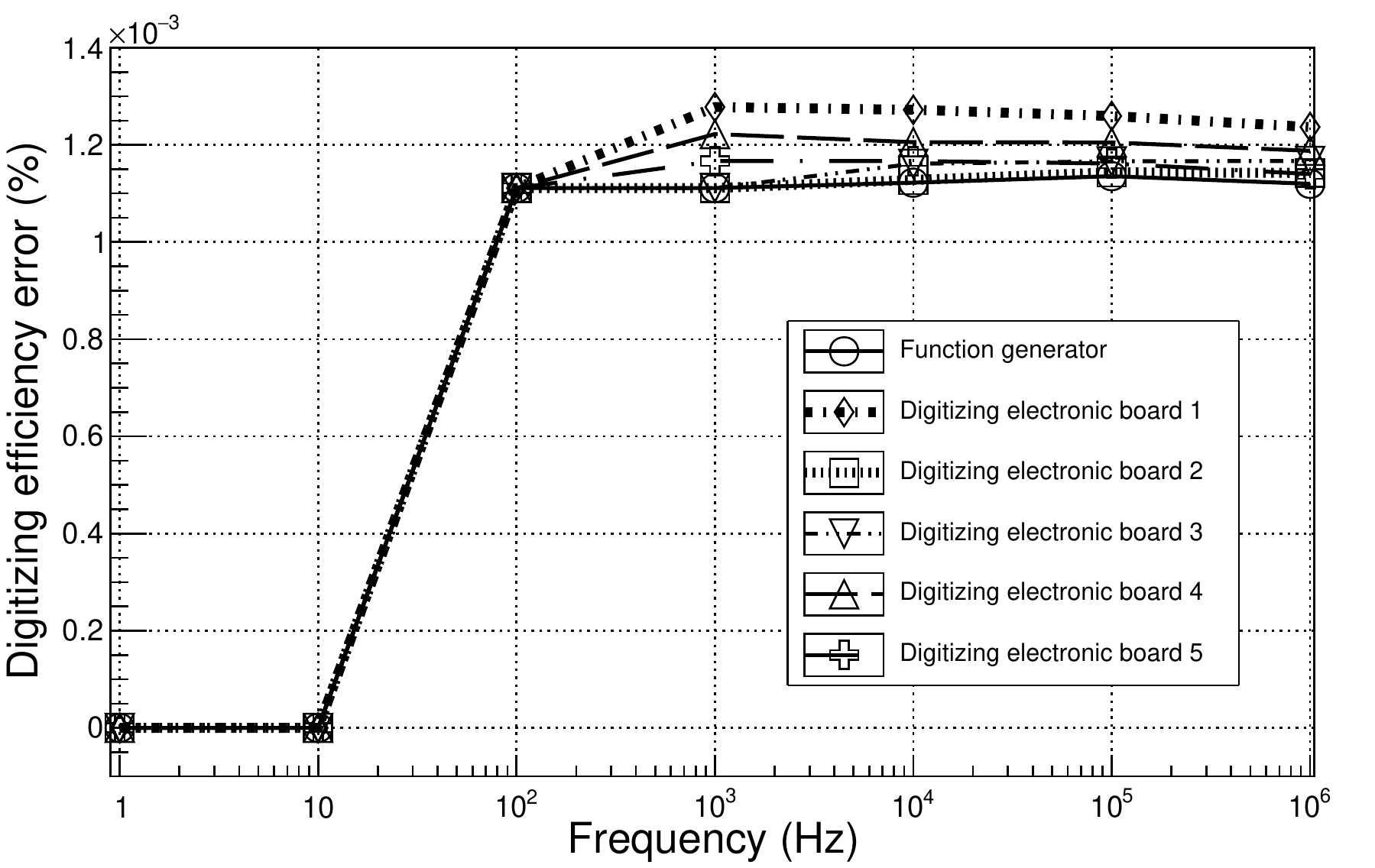}
    \caption{Digitizing efficiency error (\%) vs frequency of five digitizing electronic boards and the AFG3101 Tektronix function generator, empty circle.}
    \label{FigEFF}
\end{figure}

In \textcolor{blue}{Table \ref{table:DigitizingEfficiencyError}} and \textcolor{blue}{Fig. \ref{FigEFF}}, we show the $eff$ for five digitizing electronic boards and AFG3101 Tektronix function generator, as function of frequency.

\begin{table}[ht!]
\caption{Digitizing efficiency error (\%) for five digitizing electronic boards and AFG3101 Tektronix function generator, as function of frequency.}
\centering
\begin{tabular}{ccccccccc}
\hline\hline                      
\begin{tabular}[c]{@{}c@{}}Board\\ number\end{tabular} & 
\begin{tabular}[c]{@{}c@{}}1 Hz\end{tabular} & 
\begin{tabular}[c]{@{}c@{}}10 Hz\end{tabular} &
\begin{tabular}[c]{@{}c@{}}100 Hz\\($\times 10^{-3}$)\end{tabular} &
\begin{tabular}[c]{@{}c@{}}1 kHz\\($\times 10^{-3}$)\end{tabular} &
\begin{tabular}[c]{@{}c@{}}10 kHz\\($\times 10^{-3}$)\end{tabular} &
\begin{tabular}[c]{@{}c@{}}100 kHz\\($\times 10^{-3}$)\end{tabular} &
\begin{tabular}[c]{@{}c@{}}1 MHz\\($\times 10^{-3}$)\end{tabular} & \\
\hline	  
1 & 0.00 & 0.00 & 1.11 & 1.28 & 1.27 & 1.26 & 1.24\\
2 & 0.00 & 0.00 & 1.11 & 1.11 & 1.13 & 1.14 & 1.14\\
3 & 0.00 & 0.00 & 1.11 & 1.11 & 1.16 & 1.17 & 1.17\\
4 & 0.00 & 0.00 & 1.11 & 1.22 & 1.21 & 1.21 & 1.19\\
5 & 0.00 & 0.00 & 1.11 & 1.17 & 1.17 & 1.16 & 1.14\\
\hline 

\begin{tabular}[c]{@{}c@{}}Function\\ generator\end{tabular} & 
\begin{tabular}[c]{@{}c@{}}0.00\end{tabular} & 
\begin{tabular}[c]{@{}c@{}}0.00\end{tabular} &
\begin{tabular}[c]{@{}c@{}}1.11\end{tabular} &
\begin{tabular}[c]{@{}c@{}}1.11\end{tabular} &
\begin{tabular}[c]{@{}c@{}}1.12\end{tabular} &
\begin{tabular}[c]{@{}c@{}}1.14\end{tabular} &
\begin{tabular}[c]{@{}c@{}}1.12\end{tabular} & \\
\hline
\hline                     
\end{tabular}
\label{table:DigitizingEfficiencyError}
\end{table}

\subsubsection{Results from digitizing-electronic-board digitizing efficiency}
In this subsection, we show the results of digitizing efficiency (\%) from five digitizing electronic boards and the AFG3101 Tektronix function generator as function of the input frequency. These are some observed results: At frequencies between 1 Hz to 10 Hz, the digitizing efficiency is 100\%; at frequencies between 100 Hz to 1 MHz, the digitizing efficiency decreases; the digitizing electronic board number 2 showed the highest digitizing efficiency and very close to AFG3101 Tektronix function generator digitizing efficiency; the digitizing electronic board number 1 has lowest digitizing efficiency.

\begin{figure}[ht!]
    \centering
    \includegraphics[width=400pt]{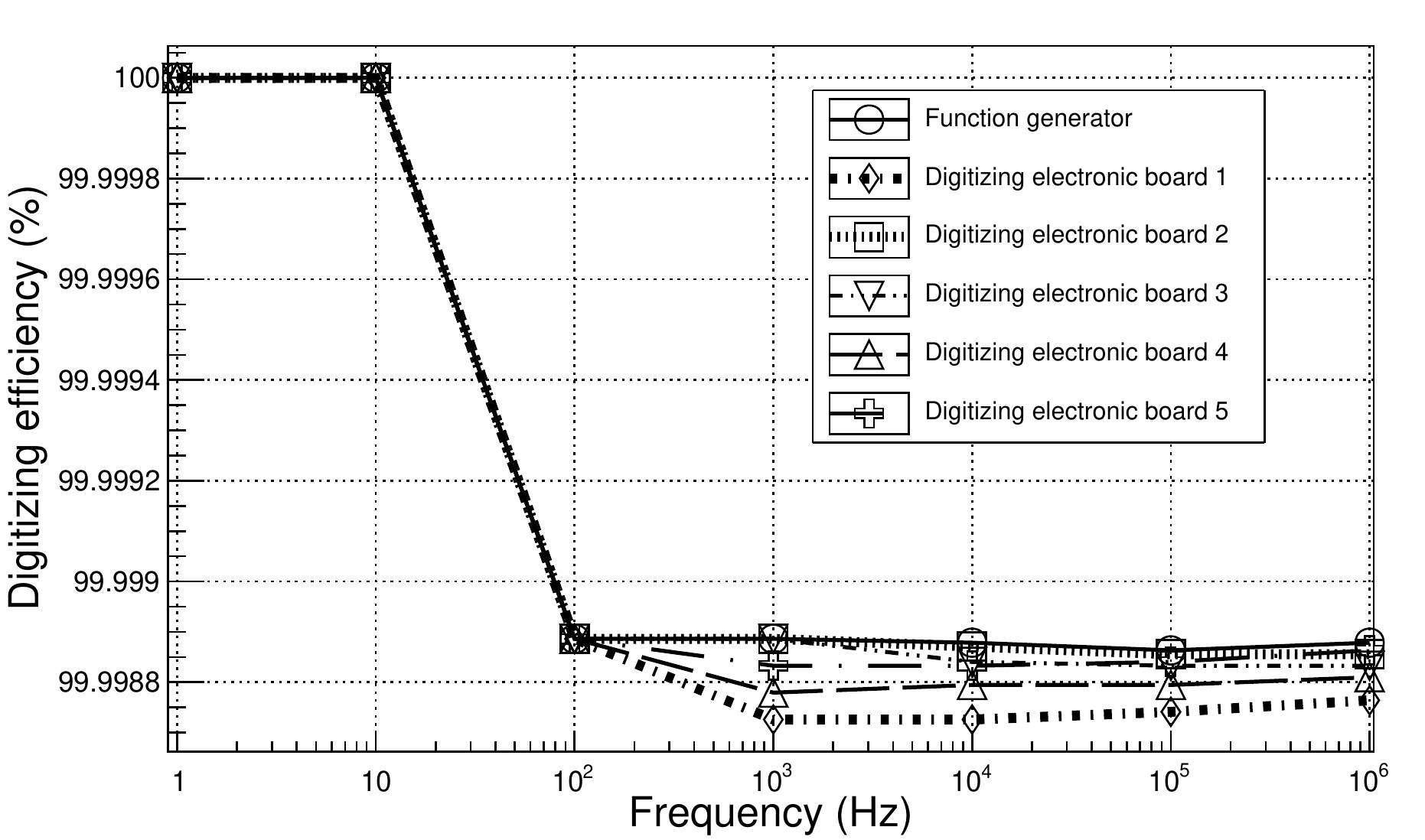}
    \caption{Digitizing efficiency of five digitizing electronic boards and AFG3101 Tektronix function generator, empty circle.}
    \label{FigDE}
\end{figure}

In \textcolor{blue}{Table \ref{table:DigitizingEfficiency}} and \textcolor{blue}{Fig. \ref{FigDE}}, we show the digitizing efficiency for five digitizing electronic boards and AFG3101 Tektronix function generator, as function of frequency.

\begin{table}[ht!]
\caption{Digitizing efficiency (\%) for five digitizing electronic boards and the AFG3101 Tektronix function generator, as function of frequency.}
\centering
\begin{tabular}{ccccccccc}
\hline\hline                      
\begin{tabular}[c]{@{}c@{}}Board\\ number\end{tabular} & 
\begin{tabular}[c]{@{}c@{}}1 Hz\end{tabular} & 
\begin{tabular}[c]{@{}c@{}}10 Hz\end{tabular} &
\begin{tabular}[c]{@{}c@{}}100 Hz\end{tabular} &
\begin{tabular}[c]{@{}c@{}}1 kHz\end{tabular} &
\begin{tabular}[c]{@{}c@{}}10 kHz\end{tabular} &
\begin{tabular}[c]{@{}c@{}}100 kHz\end{tabular} &
\begin{tabular}[c]{@{}c@{}}1 MHz\end{tabular} & \\
\hline	  
1 & 100.00 & 100.00 & 99.99889 & 99.99873 & 99.99873 & 99.99874 & 99.99876\\
2 & 100.00 & 100.00 & 99.99889 & 99.99889 & 99.99887 & 99.99886 & 99.99886\\
3 & 100.00 & 100.00 & 99.99889 & 99.99889 & 99.99884 & 99.99883 & 99.99883\\
4 & 100.00 & 100.00 & 99.99889 & 99.99878 & 99.99879 & 99.99879 & 99.99881\\
5 & 100.00 & 100.00 & 99.99889 & 99.99883 & 99.99883 & 99.99884 & 99.99886\\
\hline 

\begin{tabular}[c]{@{}c@{}}Function\\ Generator\end{tabular} & 
\begin{tabular}[c]{@{}c@{}}100.00\end{tabular} & 
\begin{tabular}[c]{@{}c@{}}100.00\end{tabular} &
\begin{tabular}[c]{@{}c@{}}99.99889\end{tabular} &
\begin{tabular}[c]{@{}c@{}}99.99889\end{tabular} &
\begin{tabular}[c]{@{}c@{}}99.99888\end{tabular} &
\begin{tabular}[c]{@{}c@{}}99.99886\end{tabular} &
\begin{tabular}[c]{@{}c@{}}99.99888\end{tabular} & \\
\hline
\hline                     
\end{tabular}
\label{table:DigitizingEfficiency}
\end{table} 

From \textcolor{blue}{Table \ref{table:DigitizingEfficiency}}, practically, the digitizing efficiency is 100\%; there are some small fluctuations.

The digitizing efficiency error and digitizing efficiency depend on the hysteresis control variable resistor (R1) of each digitizing electronic boards. 
We have two possible explanations why digitizing efficiency decreases to frequencies greater than 10 Hz in five electronic boards and AFG3101 Tektronix function generator. First, the AFG3101 Tektronix function generator and the cRIO system have a time-base of 10 MHz and 40 MHz, respectively; if we use the same frequency in their internal time-base of each instrument, the fluctuations could vanish. Second, as the phase shift of time-base of both instruments are different, if we use the external time-bases of the AFG3101 Tektronix function generator and the cRIO system and apply a Rubidium or Cesium clock as external frequency references, the fluctuations could vanish; however the cRIO system may not function properly, as 40 MHz was defined as internal time-base minimum frequency by National Instruments.

\subsubsection{Results from digitizing electronic board digitizing time}
\begin{figure}[ht!]
    \centering
    \includegraphics[width=468pt]{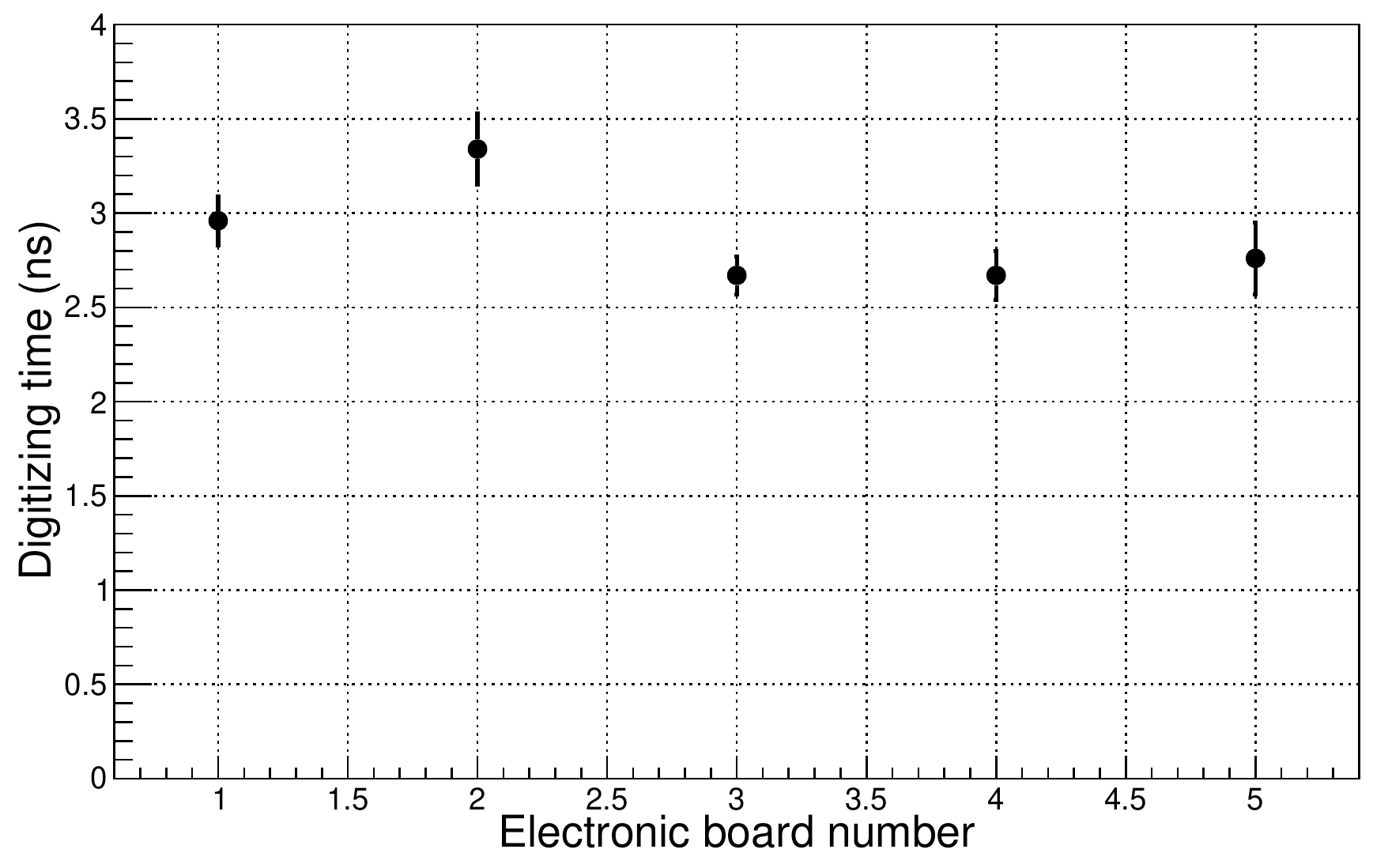}
	  \caption{Digitizing time vs electronic board number, from \textcolor{blue}{Table \ref{table:DTime}}; statistical errors only.}
    \label{FigDTime}
\end{figure}

We show subsection \textcolor{blue}{\ref{DTimeC}} results of the time the digitizing electronic boards take to convert analogue signal to digital signal. 
In \textcolor{blue}{Table \ref{table:DTime}} and \textcolor{blue}{Fig. \ref{FigDTime}}, we show the digitizing time and its error for five digitizing electronic boards. The overall digitizing time is approximately 2.88$\pm$0.15 ns; it is an average of five digitizing electronic boards digitizing time from \textcolor{blue}{Table \ref{table:DTime}}.
 
\begin{table}[!hbp]
\caption{Digitizing time for five digitizing electronic boards.}
\centering
\begin{tabular}{c c c}
\hline\hline                      
\begin{tabular}[c]{@{}c@{}}Board\\ number\end{tabular} & 
\begin{tabular}[c]{@{}c@{}}Digitizing time\\ (ns)\end{tabular} & \\
\hline          
1 & 2.96 $\pm$0.13 \\
2 & 3.34 $\pm$0.19 \\
3 & 2.67 $\pm$0.10 \\
4 & 2.67 $\pm$0.13 \\
5 & 2.76 $\pm$0.19 \\
\hline 
\hline                     
\end{tabular}
\label{table:DTime}
\end{table}

Our result is too close to the value reported by the manufacturers; the electronic channel of the digitizing electronic board consists of the ADCMP582 comparator chip (U1) \textcolor{blue}{\cite{Datasheet-ADCMP582}} and MC10ELT21DG chip (U2) \textcolor{blue}{\cite{Datasheet-MC10ELT21DG}}, see STAGE 1 of \textcolor{blue}{Fig. \ref{FigSchDiagDigitElecB}}; the digitizing time is achieved by the sum of time of 180 ps -ADCMP582 comparator chip (U1)- and 3.5 ns -MC10ELT21DG chip (U2)-.

\subsection{Results from experimental setup}
\begin{figure}[ht!]
    \centering
    \includegraphics[width=360pt]{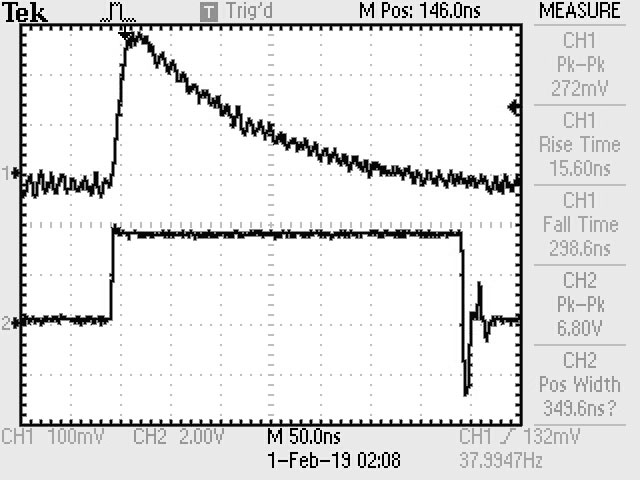}
    \caption{Waveform screenshot from experimental setup using threshold voltage at +60 mVdc and feeding voltage at +75 Vdc; analogue signal is represented by channel 1 (CH1), and digital signal is represented by channel 2 (CH2).}
    \label{FigWaveScr60mV}
\end{figure}

\begin{figure}[ht!]
    \centering
    \includegraphics[width=360pt]{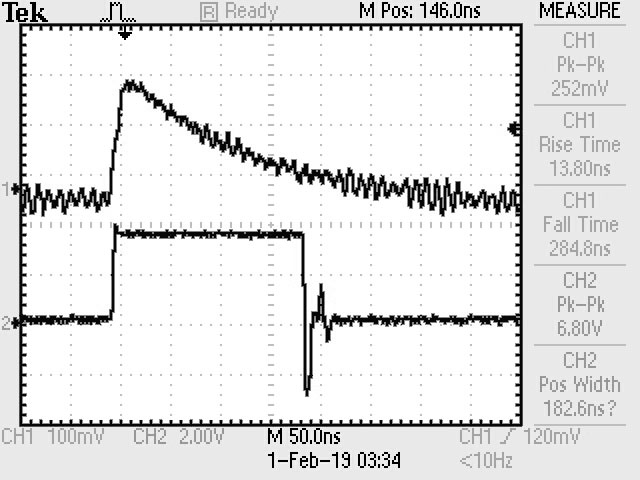}
    \caption{Waveform screenshot from experimental setup using threshold voltage at +110 mVdc and feeding voltage at +75 Vdc; analogue signal is represented by channel 1 (CH1), and digital signal is represented by channel 2 (CH2).}
    \label{FigWaveScr110mV}
\end{figure}

\begin{figure}[ht!]
    \centering
    \includegraphics[width=360pt]{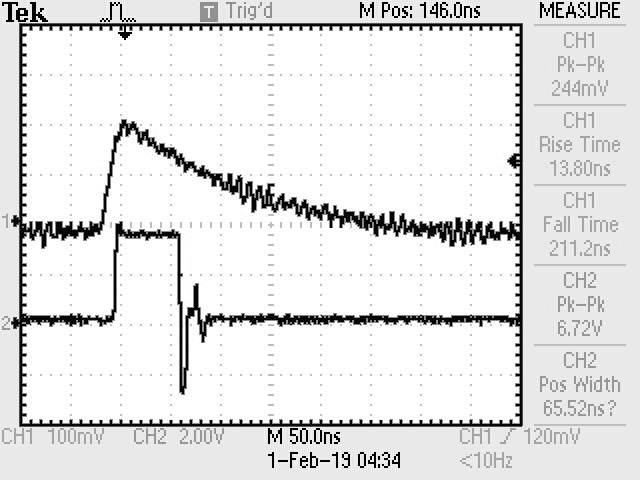}
    \caption{Waveform screenshot from experimental setup using threshold voltage at +200 mVdc and feeding voltage at +75 mVdc; analogue signal is represented by channel 1 (CH1), and digital signal is represented by channel 2 (CH2).}
    \label{FigWaveScr200mV}
\end{figure}

In this subsection we present the average basic characteristics results of the analogue signal from S12572-100P Hamamatsu photodiode and of the respective digital signal from digitizing electronic board, obtained using at least 20 waveform screenshots for each measurement, see section \textcolor{blue}{\ref{ExpSetup}}. 
In \textcolor{blue}{Table \ref{table:ThresholdVoltage}} we show an example of three average waveform screenshot samples. They correspond to +60 mVdc, +110 mVdc, and +200 mVdc of digitizing electronic board threshold voltage, and feeding voltage of +75 Vdc, see \textcolor{blue}{Fig. \ref{FigWaveScr60mV}}, \textcolor{blue}{Fig. \ref{FigWaveScr110mV}}, and \textcolor{blue}{Fig. \ref{FigWaveScr200mV}}, respectively; we show signal amplitude (V\textsubscript{Pk-Pk}), rise time (ns), and fall time (ns) of analogue signal; and we show signal amplitude (V\textsubscript{Pk-Pk}), and positive width (ns) of digital signal.

\begin{table}[!hbp]
\caption{Analogue signal and digital signal measurements detected by the S12572-100P Hamamatsu photodiode and converted by the digitizing electronic board, respectively; threshold voltage is applied to the digitizing electronic board.}
\centering
\begin{tabular}{r|ccc|crcc}
\hline\hline
\multicolumn{1}{ c }{} & \multicolumn{3}{ |c }{Analogue signal} & \multicolumn{2}{ |c }{Digital signal}\\
\begin{tabular}[c]{@{}c@{}}Threshold\\ voltage \\ (mVdc) \end{tabular} & 
\begin{tabular}[c]{@{}c@{}}Amplitude\\ (mV\textsubscript{Pk-Pk})\end{tabular} & 
\begin{tabular}[c]{@{}c@{}}Rise\\ time \\ (ns)\end{tabular} & 
\begin{tabular}[c]{@{}c@{}}Fall\\ time \\ (ns)\end{tabular} &
\begin{tabular}[c]{@{}c@{}}Amplitude\\ (V\textsubscript{Pk-Pk}) \end{tabular} & 
\begin{tabular}[c]{@{}c@{}}Positive\\ width\\ (ns)\end{tabular} & \\
\hline
 +60\;\;\;\;\; & 277 & 15.43 & 259.40 & 6.72 & 344.10\;\;\\
+110\;\;\;\;\; & 258 & 15.79 & 250.00 & 6.71 & 181.90\;\;\\
+200\;\;\;\;\; & 267 & 15.55 & 242.40 & 6.68 & 70.90\;\;\\
\hline
\hline   	
\end{tabular}
\label{table:ThresholdVoltage}
\end{table}

The experimental setup output is an  analogue signal, positive exponential decay, with amplitude of 267.33 mV\textsubscript{Pk-Pk}, rise time of 15.59 ns, and a fall time of 250.60 ns; corresponding digital signal has a 6.70 V\textsubscript{Pk-Pk} of amplitude (we detected positive overshoot and negative overshoot, by discriminating the over-shooting in the digital signal, the amplitude is very close to +5 Vdc or TTL logic family), a positive width depends of the selected threshold voltage; if the threshold voltage is near 0 Vdc, then the positive width of the digital pulse increases in time, otherwise, it decreases. The reported analogue signal and digital signal measurements are the results of the average in \textcolor{blue}{Table \ref{table:ThresholdVoltage}}.

The experimental setup noise signal result was $\sim$20 mV\textsubscript{Pk-Pk} (see step eight from subsection \textcolor{blue}{\ref{ExpSetOperation}} operations).

From the analogue signal, we observed an offset of -8 mVdc approximately; this offset is due to the connection between the feeding and reading out electronic board and digitizing electronic board; particularly, due to the installation of the 10-kohm resistor (J3) on the digitizing electronic board; the 10-kohm resistor (J3) directly affects the electronic circuit STAGE 1 from the feeding and reading out electronic board; if we remove the 10-kohm resistor (J3) of the digitizing electronic board, the offset in the analogue signal is decremented to -1 Vdc; if we decrease the 10-kohm (J3) resistor value from the digitizing electronic board, the analogue signal will present the highest attenuation; in the digital signal, an offset voltage was not observed.
From the digitizing electronic board, we observed a maximum overshoot of +2 Vdc and minimal overshoot of -3 Vdc; this overshoot doesn't affect the cRIO system measurements, because, the positive overshoot is smaller than +5 Vdc (TTL logic family), see digital signal of \textcolor{blue}{Fig. \ref{FigWaveScr60mV}}, \textcolor{blue}{Fig. \ref{FigWaveScr110mV}}, and \textcolor{blue}{Fig. \ref{FigWaveScr200mV}}, the negative overshoot is discriminated by the NI‑9402 \textcolor{blue}{\cite{Datasheet-cRIO-9402}} C module from cRIO system; the digitizing-electronic-board digitizing efficiency error is not affected by overshooting.

\section{Conclusions}
We have planned, designed, manufactured, tested and run a set of 15 electronic boards (ten feeding and reading out electronic boards and five digitizing electronic boards) to connect, to electrically feed the S12572-100P Hamamatsu photodiode, to read out its signals, and to digitize them. All electronic boards, four front-end electronic boards' characterization, one experimental setup, and cRIO DAQ system worked properly.

From the experimental setup used to evaluate the S12572-100P Hamamatsu photodiode, the digital signal corresponds perfectly to the analogue signal reported by S12572-100P Hamamatsu photodiode.

We have evaluated attenuation factor, phase shift, and transit time of ten feeding and reading out electronic boards; the results were similar between them. Attenuation factor was measured to be $\sim$0.80; phase shift, at average very low, but at 400 kHz and 800 kHz, the phase shift was big; transit time, $\sim$738.08$\pm$62.38 ps.

We have evaluated digitizing efficiency error, digitizing efficiency, and digitizing time of five digitizing electronic boards; the results are similar between them for both digitizing efficiency error and digitizing efficiency. Digitizing efficiency error was measured to be $\sim$1.11$\times 10^{-3}$\%; digitizing efficiency, $\sim$99.99\% value typically; digitizing time, 2.88$\pm$0.15 ns.

From the above measurements, the proposed electronic circuit in electronic boards is very convenient to electrically feed the S12572-100P Hamamatsu photodiode, to read out its analogue signals and to digitize them. The applications of this photodiode are very far-reaching. The costs of operations, compared with photomultiplier, are substantially reduced.

\section*{Supplementary material}
The supplementary material is divided into five compressed folders, as follows:

\begin{enumerate}[1.]
\item The Characterization-1.zip folder contains attenuation factor and phase shift material. This supplementary material includes a LabVIEW program (VI program developed by us), ten data text file supplemental results, ten attenuation factor plots (the first 100 measurements), ten attenuation factor plots (the remaining measurements), ten attenuation factor plots (199 measurements), ten linear scale phase shift plots (199 measurements), ten logarithmic scale phase shift plots (199 measurements), and five scripts in C code.

\item The Characterization-2.zip folder contains digitizing efficiency error material and digitizing efficiency material. This supplementary material includes a cRIO-9025 LabVIEW project\footnote{This project can be compiled on any National Instruments cRIO controller.} (project and VI's developed by us), 42 data text files supplemental results, two plots, and one script in C code.

\item The Characterization-3.zip folder contains transit time material. This supplementary material includes at least 30 screenshots of a waveform for each feeding and reading out electronic boards, ten data text file supplemental results, one plot, one LabVIEW program, and one script in C code.

\item The Characterization-4.zip folder contains digitizing time material. This supplementary material includes at least 20 screenshots of a waveform for each digitizing electronic boards, five data text file supplemental results, one plot, one LabVIEW program, and one script in C code.

\item The ExperimentalSetup.zip folder includes at least 29 screenshots of a waveform corresponding to +60 mVdc of threshold voltage, at least 26 screenshots of a waveform corresponding to +110 mVdc of threshold voltage, at least 32 screenshots of a waveform corresponding to +200 mVdc of threshold voltage, three data text file supplemental results, and one script in C code.

\item The References.zip folder includes datasheets, user manuals, and programmer manuals of the electronic circuits, connectors, cables, power supplies, Tektronix oscilloscopes, and Tektronix function generator used on this reseach.
\end{enumerate}

\href{https://drive.google.com/drive/folders/1JV1KHiFaHEMyfcZ1JRZ0864kKgq07PoG?usp=sharing}{\textcolor{blue}{\underline{Click here}}} for all the supplementary material.

\section*{Acknowledgements}
The authors gratefully acknowledge the support of Universidad de Guanajuato, Guanajuato, M\'{e}xico.

The corresponding author wishes to thank Roberto Valdivia (student of Departamento de F\'{i}sica de la Divisi\'{o}n de Ciencias e Ingenier\'{i}as campus Le\'{o}n – Universidad de Guanajuato) for his collaboration in obtaining the photographic and video material, and for co-producing the first version of three video presentation.

The corresponding author wishes to thank Juan Arceo for producing and narrating the three video presentations for this publication.

\raggedright
\begin{sloppypar}
\bibliography{mybibfile}
\end{sloppypar}
\end{document}